\documentclass[a4paper,11pt]{article}
\pdfoutput=1 

\usepackage{jheppub} 

\usepackage[T1]{fontenc} 

\title{Weyl holographic superconductor in the Lifshitz black hole background}

\author[a,b]{S. A. Hosseini Mansoori,}
\author[b]{B. Mirza,}
\author[c]{A. Mokhtari,}
\author[b]{F. Lalehgani Dezaki,}
\author[b]{Z. Sherkatghanad}

\affiliation[a]{Department of Physics, Boston University, 590 Commonwealth Ave., Boston, MA 02215, USA}
\affiliation[b]{Department of Physics, Isfahan University of Technology, Isfahan 84156-83111, Iran}
\affiliation[c]{Department of Physics, Tarbiat Modares University, Tehran 14155-4838, Iran}

\emailAdd{shossein@bu.edu}
\emailAdd{b.mirza@cc.iut.ac.ir}
\emailAdd{ali.mokhtari@modares.ac.ir}
\emailAdd{f.Lalehgani@ph.iut.ac.ir}
\emailAdd{sherkat.elham@gmail.com}

\abstract{ We  investigate analytically the properties of the Weyl holographic superconductor in the
Lifshitz black hole background. We find that the critical temperature
of the Weyl superconductor decreases with increasing  Lifshitz dynamical exponent, $z$, indicating that condensation becomes
difficult. In addition, it is found that the critical temperature and condensation operator could be affected by applying the Weyl coupling, $\gamma$. Moreover, we compute the critical magnetic field and investigate its dependence on the parameters $\gamma$ and  $z$. Finally, we show numerically that the Weyl coupling parameter $\gamma$ and the Lifshitz dynamical exponent $z$ together control the size and strength of the conductivity peak and the ratio of gap frequency over critical temperature $\omega_{g}/T_{c}$.}

\begin{document} 
\maketitle
\flushbottom

\section{Introduction}\label{se1}
Superconductivity is one of the most important phenomena in condensed matter physics
(CMP) which is characterized by the drop of  electrical resistivity to zero at a critical temperature $T_{c}$ and an expulsion of the magnetic field from the interior of a sample ~\cite{Mref2, Mref1}.

In 1957, Bardeen, Cooper, and Schrieffer published two articles which established the conceptual and mathematical foundations of conventional superconductivity, for which they later received their Nobel Prize in 1972 ~\cite{Jref1, Jref2}.
According to the BCS theory, the condensate is a Cooper pair of electrons bounded together by phonons. Therefore, the main effective coupling in the BCS theory is the electron-phonon coupling which is considered to be  a weak coupling constant.
High-temperature medium-coupled superconductivity predicted for hydrides at high pressures indicates that high-temperature phonon-mediated superconductivity can also be described by the BCS theory ~\cite{Struzhkin201577,Mahdi}. However, there are certain materials that exhibit unconventional superconductivity at high $T_{c}$ cuprates and thus, their explanation calls for  new theories to be developed. It is by now clear that the models for high-temperature superconductivity must be formulated as theories in the strong coupling constant regime.

The Anti-de Sitter/Conformal Field Theory (AdS/CFT) correspondence is one  way to describe  the strong coupling constant regime ~\cite{reff1, reff2} and the AdS/CFT duality, indeed, provides  us with a new theoretical framework to realize the physics of high $T_{c}$ superconductors. This duality, based on the holographic principle, establishes a relationship between gravitational theories on AdS spacetime and a quantum field theory that lives on the conformal boundary of the AdS spacetime.
Applying the AdS/CFT to unconventional superconductors allows us achieve a dual gravitational description of a superconductor involving the mechanism of spontaneous breaking of Abelian gauge symmetry near the event horizon of the black hole, which leads to the formation of a scalar hair condensate at a temperature $T$ less than a critical temperature $T_{c}$ ~\cite{Dref1, Dref2, Dref3, Dref4, Dref5,Dref6,Dref7}. This concept leads to a major advance in the study of holographic superconductors, which strongly depends on the properties of AdS black holes. The presence of the AdS black holes with a dual role in the holographic superconductor model  provides a non-zero temperature for the boundary relativistic CFT and forms a scalar hair condensation at the boundary. However, the real condensed matter systems are far from a relativistic one. Therefore, it is interesting to generalize these holographic superconducting models to non-relativistic situations ~\cite{Sref11, Sref1, Sref2, Sref3, ref5,refff1,refff2}. 

A generalization of the AdS/CFT correspondence to non-relativistic conformal field theory (NR-CFT) was investigated for the first time in Ref. \cite{Refre1}. The NR-CFT is invariant under Galilean transformations with Schr\"{o}dinger symmetry for systems govern ultra cold atoms at unitarity, nucleon scattering, and family of universality classes of quantum critical behavior. Among other novel results, it was further argued in Ref. \cite{Refre2} that non-relativistic CFT that describe multicritical points in certain magnetic materials and liquid crystals may be dual to certain non-relativistic gravitational theories in the Lifshitz space-time background.

Recently, Horava proposed a non-relativistic renormalizable theory of gravitation, the so-called Horava-Lifshitz (HL) theory, which reduces to Einstein's general relativity at large distances \cite{Refre3}. Additionally, the HL theory may render a candidate for a UV completion of Einstein's theory. Also at short distances, the spacetime manifold is equipped with an extra structure, of a fixed codimension-one foliation by slices of constant time which defines a global causal structure \cite{Refre3, Refre4}. Moreover, HL gravity provides the minimal holographic dual for Lifshitz-type field theories with anisotropic scaling ($t \to \lambda^{z}t$, $x \to \lambda x$) and dynamical exponent $z$ \cite{Refre5}.  In Ref \cite{Refre5} it was also shown that the Lifshitz spacetime is a vacuum solution of the HL gravity.
Meanwhile, another form of nonrelativistic holography with HL gravity were also presented in \cite{Refre6,Refre7}. In these articles, authors investigated a non-relativistic gravity theory (HL gravity) dual for any NR-CFTs which have the same set of symmetry transformations such as time dependent spatial diffeomorphisms, spatially dependent temporal diffeomorphisms, and the $U(1)$ symmetry acting on the background gauge field coupled to particle number. Specially, time dependent spatial diffeomorphisms include the Galilean invariant. Therefore, both approaches in Refs. \cite{Refre5} and \cite{Refre6,Refre7} of non-relativistic holography demonstrate that the natural arena for non-relativistic holography is non-relativistic HL gravity. 

On the other hand, in Ref. \cite{Refre8}, it was shown that the IR action of the non-projectable HL gravity exhibit asymptotically Lifshitz black hole. The main thrust of the above discussion is that  Lifshitz black hole background inherently satisfies Galilean invariant.
Moreover, from a holographic point of view, the Lifshitz black holes  capture the non-relativistic behavior at finite temperatures for the boundary CFT ~\cite{ref4}. The metric of the Lifshitz black holes reproduce asymptotically the Lifshitz spacetime \cite{ref4, ref441}.

In what follows, a holographic superconductor model is constructed by making use of Lifshitz black hole solutions and  the effects of the dynamical critical exponent, $z$, are investigated on the properties of such holographic superconductors. To date, holographic superconducting models have been generalized in a number of studies to non-relativistic situations by Lifshitz black hole backgrounds ~\cite{Sref1, Sref2, Sref3}. In our symmetry broken phase, there is a Goldstone which may lead to  nonzero conductivity at low frequencies. The real part of conductivity, however, vanishes in these simple holographic  models at zero temperature and at low frequencies ~\cite{Sref1, Sref2, Sref3}. In this paper, we, therefore, consider a dual gravitational description of a holographic superconductor using a particular form of unprecedented higher derivative corrections, which involves couplings between the gauge field and the spacetime curvature of the Lifshitz black hole. More precisely, we assume an action for the bulk Abelian gauge field contains the Maxwell term as well as the coupling between the Weyl tensor and the field strengths. The Weyl coupling, $\gamma$, is usually constrained by respecting the causality of the dual field theory on the boundary and by preserving the positivity of the energy flux in the CFT analysis ~\cite{ref3, R. C. Myers, A. Ritz, ref7}.

 For our purposes, the  constraints imposed on the Weyl coupling in the Lifshitz background and in $d$ dimensions will be initially explored by demanding that the dual CFT should respect the causality and that  the energy flux be positive in all directions for the boundary CFT. Our results reduces to the one obtained in Ref. ~\cite{R. C. Myers} when the dynamical exponent is $z = 1$ in four dimensions. Moreover, these results indicate that the upper bound of the Weyl coupling in Ref. ~\cite{A. Ritz} is modified by imposing the positivity condition on the energy flux for the boundary CFT. We investigate the effects of the Lifshitz dynamical exponent, $z$, and the Weyl coupling, $\gamma$, on the holographic superconductors in the Lifshitz black hole background and in $d$ dimensions, assuming that the back reaction effects are negligible and take the probe limit. For $3+1$ dimensions and  $z=1$  Ref. ~\cite{ref6}  may be consulted. We use the matching method of the field solutions, near the horizon and the boundary, to study the response of a holographic superconductor to an external magnetic field in the presence of Weyl corrections. The analysis reveals that the critical magnetic field is affected by both the Lifshitz dynamical exponent and the Weyl coupling parameter. We would also like to investigate numerically whether the universal relation between the gap $\omega$ in the frequency dependent conductivity and the critical temperature $T_{c}$, $\omega_{g}/T_{c} \approx 8$ obtained in \cite{Dref1}, is stable under Weyl corrections in Lifshitz black hole background.

The paper is organized as follows. In Section ~\ref{cap2}, we obtain the explicit constraints on the Weyl coupling that respect the causality of the dual field theory on the boundary and preserve  the positivity of the energy flux in the CFT analysis. We show that the upper bound  of the Weyl coupling in Ref. ~\cite{A. Ritz} is modified. The holographic superconductor is then constructed using the  Weyl corrections and the effects of $z$ and $\gamma$ are investigated on the condensation and critical temperature of the holographic superconductor. Section  ~\ref{cap3} is devoted to the study of the properties of the Weyl holographic superconductors with Lifshitz scaling in the presence of an external magnetic field. Then we move on to investigate the electromagnetic fluctuations of the system and numerically calculate the electrical conductivity using linear response theory in Section ~\ref{cap5}. Finally,  conclusions are presented in Section ~\ref{cap4}.

\section{Weyl coupling and the condensate operator}\label{cap2}
\begin{figure}[tbp]
\centering
\includegraphics[scale=.5]{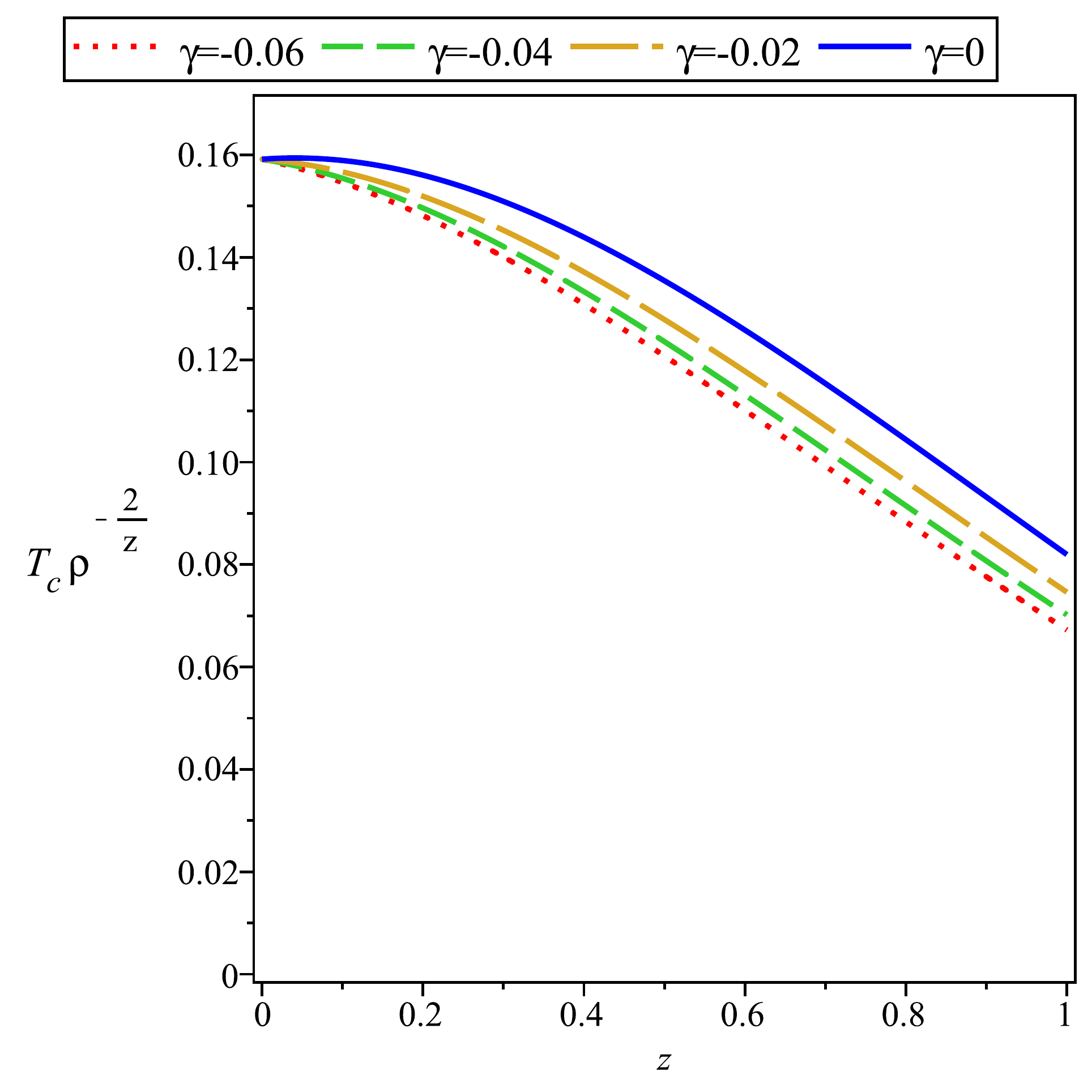}
\caption{\label{pic1}Value of  critical temperature as a function of  dynamical exponent, $z$, for the following parametric values  $d=2$, $u_{i}=0.5$, and
$m^2=-0.5$.  }
\end{figure}

 This Section begins with introducing the Lifshitz background before  the Weyl superconductor  is investigated.  The Liftshitz black hole with the flat horizon takes the following form ~\cite{ref4}:
 \begin{equation}\label{q1}
d{{s}^{2}}={{L}^{2}}\left( -{{r}^{2z}}f\left( r \right)d{{t}^{2}}+\frac{d{{r}^{2}}}{{{r}^{2}}f\left( r \right)}+{{r}^{2}}\sum\limits_{i=1}^{d}{d{{x}_{i}}^{2}} \right)
\end{equation}
with
\begin{equation}\label{q2}
f(r)=1-{{\left( \frac{r_{0}}{r} \right)}^{d+z}}\,\,\  ; \,\,\ {{r}_{0}}\le r<~\infty
\end{equation}
where, $r_{0}$ and $z$ are the black hole horizon and the dynamical exponent, respectively.
It is easy to show that the Liftshitz background has the following  Lifshitz scaling symmetry at the Liftshitz spacetime when $f(r) = 1$,
\begin{equation}
t\to {{\lambda }^{z}}t \,\,\ ; \,\,\ r\to \frac{r}{\lambda }\,\,\ ; \,\,\ \vec{x}\to \lambda \vec{x}
\end{equation}
The Liftshitz metric \eqref{q1} also reduces to the Schwarzschild-$AdS_{d+2}$ black hole metric when
$z = 1$. Under the transformation $u={r_{0}}/r$, the metric \eqref{q1} can also be recast  in the   following form:
\begin{eqnarray}\label{q4}
 d{{s}^{2}}={{L}^{2}}\left( \frac{-f\left( u \right) r_{0}^{2z}}{{{u}^{-2z}}}d{{t}^{2}}+\frac{d{{u}^{2}}}{{{u}^{2}}f\left( u \right)}+\frac{r_{0}^{2z}}{{{u}^{2}}}\sum\limits_{i=1}^{d}{d{{x}^{2}}} \right)
\end{eqnarray}
where, $f\left( u \right)=1-{{u}^{d+z}}$ and the $u$ coordinate maps the holographic direction to the finite interval $\left( 0,1 \right]$. Furthermore, the Hawking temperature of the Lifshitz black hole is given by:
\begin{equation}\label{q7}
T_{H}=\frac{z+d}{4\pi L^{2} }r_{0}^{z}
\end{equation}
In the rest of this paper,  we will set $L = 1$ to simplify the calculations. We will also consider a Weyl correction of the Maxwell field and a charged complex scalar field coupled via the following Lagrangian density:
\begin{equation}\label{EQ1}
{{L}_{W}}=-\frac{1}{4}\left[ {{F}^{\mu \nu }}{{F}_{\mu \nu }}-4\gamma {{C}^{\mu \nu \rho \sigma }}{{F}_{\mu \nu }}{{F}_{\rho \sigma }} \right]-{{\left| {{D}_{\mu }}\psi  \right|}^{2}}-{{m}^{2}}{{\left| \psi  \right|}^{2}}
\end{equation}
where, ${{D}_{\mu }}={{\nabla }_{\mu }}-i{{A}_{\mu }}$, ${{F}_{\mu \nu }}={{\nabla }_{\mu }}{{A}_{\nu }}-{{\nabla }_{\nu }}{{A}_{\mu }}$, and $m$ is the mass of the scalar field, $\psi$. Moreover,  $\gamma$ is a dimensionless constant with a limit on it while ${{C}_{\mu \nu \rho \sigma }}$ is the Weyl tensor. From this Lagrangian,  the generalized scalar and vector equations of motion may be expressed as follows:.
\begin{eqnarray}
& {{D}_{\mu }}{{D}^{\mu }}\psi -{{m}^{2}}\psi =0 \label{q5} \\
 &   {{\nabla }_{\mu }}\left[ {{F}^{\mu \nu }}-4\gamma {{C}^{\mu \nu \rho \sigma }}{{F}_{\rho \sigma }} \right]=i\left( {{\psi }^{*}}{{D}^{\nu }}\psi -\psi {{D}^{\nu *}}{{\psi }^{*}} \right)\label{q6}
\end{eqnarray}
The non-vanishing components of the Weyl tensor in the Liftshitz background may also be listed as follows:
\begin{align}
 {{C}_{tutu}}=\left( \begin{array}{c}
  d  \\
  2  \\
\end{array} \right)\frac{r_{0}^{2z}\xi (u)}{{{u}^{2z+2}}} \,\,\,\ ; \,\,\,\   {{C}_{titj}}=\left( \frac{d-1}{2} \right)\frac{r_{0}^{2z+2}f(u)\xi (u)}{{{u}^{2z+2}}}{{\delta }_{ij}} \\
\nonumber {{C}_{uiuj}}=-\left( \frac{d-1}{2} \right)\frac{r_{0}^{2}\xi (u)}{{{u}^{4}}f(u)}{{\delta }_{ij}}  \,\,\,\ ; \,\,\,\   {{C}_{ijkl}}=-\frac{r_{0}^{4}\xi (u)}{{{u}^{4}}}{{\delta }_{ij}}{{\delta }_{kl}}
\end{align}
where,
\begin{equation}
\xi \left( u \right)=\frac{z}{d+1}\left[ {{u}^{d+z}}\left( \frac{d+2}{z}-1 \right)+\frac{2}{d}(1-z) \right]
\end{equation}
Now,  the constraints imposed on the Weyl coupling, $\gamma$ as sought by demanding
that the dual CFT should respect both the causality ~\cite{ref1} and the stability of the modes for the vector field which indicates that the uniform
neutral plasma is a stable configuration in the dual CFT ~\cite{ref2, ref3}. To examine causality and the stability of the modes, the  Ginzburg-Landau terms ($\psi=0$) are initially ignored  in action \eqref{EQ1} . Therefore, Maxwell's equation can be rewritten as follows:
\begin{equation}\label{g35}
{{\nabla }_{\mu }}\left( {{F}^{\mu \nu }}-4\gamma {{C}^{\mu \nu \rho \sigma }}{{F}_{\mu \nu }} \right)=0
\end{equation}
The Fourier-space representation of the gauge field is:
\begin{equation}
{{A}_{a}}(t,x,{{y}_{i}},u)=\int{\frac{{{d}^{3}}q}{{{(2\pi )}^{3}}}{{e}^{i\mathbf{q}.\mathbf{x}}}}{{A}_{a}}(\mathbf{q},u)	
\end{equation}
where, $\mathbf{q}.\mathbf{x}=-\omega t+{{q}^{x}}x+{{q}^{{{i}}}}{{y}_{i}}$ with $i=1,2,...,d-1$. It is also convenient to select the momentum to be ${{\mathbf{q}}^{\mu }}=(\omega ,q,0,\cdots ,0)$ with $(d-1)$ zero components and the gauge field ${{A}_{u}}(q,u)=0$. Substituting this term into Eq. \eqref{g35} and  considering the Liftshitz black hole background (Eq. \eqref{q4}), we find the following expressions:
\begin{eqnarray}
0&=&{{A}_{t}}^{''}+A_{t}^{'}\left( \frac{{{U}_{1}}'}{{{U}_{1}}}+\frac{z-d+1}{u} \right)+\frac{q}{f}\frac{{{U}_{2}}}{{{U}_{1}}}\left( q{{A}_{t}}\left( u \right)+w{{A}_{x}}\left( u \right) \right)\label{g1}\\
0&=&{{A}_{x}}^{''}+A_{x}^{'}\left( \frac{{{U}_{2}}'}{{{U}_{2}}}+\frac{f'}{f}-\frac{z+d-3}{u} \right)+\frac{w{{u}^{2z-2}}}{{{f}^{2}}}\left( q{{A}_{t}}\left( u \right)+w{{A}_{x}}\left( u \right) \right) \label{g2}\\
0&=&{{A}_{y_{i}}}^{''}+A_{y_{i}}^{'}\left( \frac{{{U}_{2}}'}{{{U}_{2}}}+\frac{f'}{f}-\frac{z+d-3}{u} \right)+{{A}_{y_{i}}}\left( u \right)\left( \frac{{{w}^{2}}}{{{f}^{2}}}{{u}^{2z-2}}-\frac{{{q}^{2}}}{f}\frac{{{U}_{3}}}{{{U}_{2}}} \right) \label{g3}\\
0&=&A_{x}^{'}-\left( \frac{{{U}_{1}}}{{{U}_{2}}}\frac{w}{q}\frac{{{u}^{2z-2}}}{f} \right)A_{t}^{'} \label{g4}
\end{eqnarray}
where, $U_{1}$, $U_{2}$, and $U_{3}$ are defined as:
\begin{eqnarray}\label{ssq1}
\nonumber {{U}_{1}}&=&\left( -\frac{1}{4}+2\gamma \left( \frac{d(d-1)}{2} \right)\xi (u) \right)\\
{{U}_{2}}&=&\left( \frac{1}{4}+2\gamma \left( \frac{d-1}{2} \right)\xi (u) \right)\\
\nonumber    {{U}_{3}}&=&\left( \frac{1}{4}-2\gamma \xi (u) \right)
\end{eqnarray}
According to the first two Eqs. \eqref{g1} and \eqref{g4}, one can decouple the equation of motion for $A_{t}(q,u)$ as follows:
\begin{equation}\label{g8}
{{{A}'''}_{t}}+{{H}_{1}}\left( u \right){{{A}''}_{t}}+{{H}_{2}}\left( u \right){{{A}'}_{t}}=0
\end{equation}
where, ${H}_{1}$ and ${H}_{2}$ are obtained as follows:
\begin{eqnarray}\label{g9}
{{H}_{1}}&=&\frac{{{z}_{1}}}{u}+\frac{{{f}'}}{f}+\frac{2{{{{U}'}}_{1}}}{{{U}_{1}}}-\frac{{{{{U}'}}_{2}}}{{{U}_{2}}}\\
\nonumber {{H}_{2}}&=&\frac{{{{{U}''}}_{1}}}{{{U}_{1}}}+\frac{{{{{U}'}}_{1}}}{{{U}_{1}}}\left[ \frac{{{z}_{1}}}{u}+\frac{{{f}'}}{f}-\frac{{{{{U}'}}_{2}}}{{{U}_{2}}} \right]+\frac{{{z}_{1}}}{u}\left[ \frac{{f}'}{f}-\frac{1}{u}-\frac{{{{{U}'}}_{2}}}{{{U}_{2}}} \right]+\frac{{{U}_{2}}{{q}^{2}}}{{{U}_{1}}f}+\frac{{{\omega }^{2}}{{u}^{2z-2}}}{{{f}^{2}}}
\end{eqnarray}
where, $z_{1}=z-d+1$. In order to investigate the causality at the CFT boundary and the stability of quasi-normal modes in the bulk theory, the full wave functions \eqref{g8} and \eqref{g3} need to be rewritten  in the form of the Schr\"{o}dinger
equation ( See Appendix ~\ref{AppA}).
By using WKB approximation in the limit $q \to \infty$, $V_{0}(u)$ (Eq. \eqref{g14}) and $W_{0}(u)$ (Eq.\eqref{g14} ) will be  the effective potentials. One can easily examine the behaviors of $V_{0}(u)$ and $W_{0}(u)$ near the
boundary, i.e., $u = 0$ ~\cite{ref1,ref2, ref3}. In order to verify the causality in the dual CFT, we need to consider the following limitations on the expansion of the effective potentials at the boundary:
\begin{equation}\label{QEQ2}
\left\{ \begin{array}{cc}
&   {{V}_{0}}\left( u\to 0 \right)< 1  \\
 &  {{W}_{0}}\left( u\to 0 \right)< 1  \\
\end{array} \right.
\end{equation}
Moreover, in the WKB limit, the potential has a minimum near the horizon ($u=1$). These effective potentials, $V_{0}(u)$ and $W_{0}$, show bound states with negative energies, which correspond to unstable quasi-normal modes in the bulk theory. For stability, we demand that the energy should be positive in all directions for a consistent CFT. For this purpose, we should consider the following limitations on the expansion of $V_{0}(u)$ and $W_{0}(u)$ near $u = 1$ ~\cite{ref2, ref3}.
\begin{equation}\label{EQQ3}
\left\{ \begin{array}{c}
   {{V}_{0}}(u\to 1)>0  \\
   {{W}_{0}}\left( u\to 1 \right)>0  \\ \end{array} \right.
\end{equation}
We, therefore, need to investigate the gamma bound in different cases. Based on the expansion of ${{W}_{0}}(u)$ and ${{V}_{0}}(u)$ potentials near the boundary and due to the causality requirement \eqref{QEQ2},  a limited range of the Weyl coupling is obtained for $0<z<1$ , $z=1$, and $z>1$  (See Appendix ~\ref{AppB}). The results show that there is no constraint on the Weyl coupling for $0<z<1$. For $z=1$, the bound is  obtained as follows:
\begin{eqnarray}\label{g23}
\gamma_{4}<\gamma <\gamma_{1}
\end{eqnarray}
Finally, we have the  bound below for $z>1$.
\begin{eqnarray}\label{g24}
 \gamma >\gamma_{2} \,\,\ and \,\,\ \gamma <\gamma_{3}
\end{eqnarray}
where, $\gamma_{1}$, $\gamma_{2}$, $\gamma_{3}$, and $\gamma_{4}$ are defined as in Appendix ~\ref{AppB}.
It should be noted  that the above bounds are the intersections of the bounds obtained from both potentials.
On the other hand, based on the large momenta limit of our effective potentials \eqref{EQQ3} and by expanding these potentials close to the horizon ( See Appendix ~\ref{AppB}), the following Weyl coupling range is again obtained  for the three situations of $d=2z-2$ and $d\ne 2z-2$.
\begin{equation}
 \begin{array}{cc}\label{g25}
   {{\gamma }_{5}}<\gamma <{{\gamma }_{6}} & for\,\ d>~2z-2  \\
   No \,\ constraint & for\,\ d=2z-2  \\
   {{\gamma }_{6}}<\gamma <{{\gamma }_{5}} & for\,\ d<2z-2  \\ \end{array}
\end{equation}
where,
\begin{equation}
{{\gamma }_{5}}=-\frac{d\left( d+1 \right)}{4\left( d-1 \right)\left( d+z \right)\left( d-2z+2 \right)} \,\,\ ; \,\,\   {{\gamma }_{6}}=\frac{d+1}{4\left( d-1 \right)\left( d+z \right)\left( d-2z+2 \right)}
\end{equation}
The above bounds are valid for $z\ne 1$ and the constraint \eqref{g23} is valid for$z=1$. We are now  in a position to find a new bound on the Weyl coupling by intersecting both bounds obtained from the effective potential expansion near the horizon and the boundary. Hence, the bound $\gamma$  for $0<z<1$ can be expressed as:
\begin{equation}
{{\gamma }_{5}}<\gamma <{{\gamma }_{6}}
\end{equation}
Moreover, it will have the following  range for $z=1$.
\begin{eqnarray}\label{g28}
\gamma_{4}<\gamma <\gamma_{1}
\end{eqnarray}
Clearly, in the case of $d=2$, the constraint on the coupling $\gamma$ ($-1/12<\gamma<1/12$) is in agreement with the result reported in Ref. ~\cite{R. C. Myers}. Moreover, in Ref. ~\cite{A. Ritz}, the authors show that the limit on the parameter $\gamma$ is $-1/16<\gamma<1/24$, where the upper bound is due to the existence of an additional singular point when $\gamma=1/24$ and the lower bound is because of the causality constraint. In our work, the gamma bound is $-1/16<\gamma<1/32$, where the upper bound is modified by considering the constraints on the effective potentials due to the stability of the modes. When $1<z$ and $d=2z-2$, the constraint on the Weyl coupling is also given by:
\begin{eqnarray}\label{g29}
\gamma >~{{\gamma }_{2}} \,\ and \,\ \gamma <{{\gamma }_{3}}
\end{eqnarray}
For $z>1$, we cannot express the explicit relation for the gamma bound as a function of $(z,d)$ for $d\ne2z-2$; thus, one needs to compute the intersection of Eqs. \eqref{g24} and \eqref{g25} for this case. For $z-2<d<2z-2$, there is an explicit formula for the gamma bound as follows:
\begin{eqnarray}\label{g29}
{{\gamma }_{2}}<\gamma <{{\gamma }_{5}}
\end{eqnarray}
Let us now  return to the Weyl holographic superconductor. Considering the following ansatz for the scalar  and Maxwell fields:
\begin{equation}
\psi =\psi (u) \,\,\ ; \,\,\ {{A}_{\mu }}d{{x}^{\mu }}=\varphi (u)dt
\end{equation}
the equations of motion \eqref{q5} and \eqref{q6} in the background \eqref{q4} reduce to:
\begin{equation}\label{E9}
{{\psi }^{''}}\left( u \right)+{{\psi }^{'}}\left( u \right)\left[ \frac{{{f}^{'}}\left( u \right)}{f\left( u \right)}-\frac{d+z-1}{u} \right]+\psi \left( u \right)\left[ \frac{{{u}^{2z-2}}{{\varphi }^{2}}\left( u \right)}{r_{0}^{2z}{{f}^{2}}\left( u \right)}-\frac{{{m}^{2}}}{{{u}^{2}}f\left( u \right)} \right]=0
\end{equation}
\begin{equation}\label{E8}
\varphi ''\left( u \right)+\varphi '\left( u \right)\left[ \left( \frac{z-d+1}{u} \right)+\frac{U{{'}_{d,z}}(u)}{{{U}_{d,z}}(u)} \right]-\frac{2{{\psi }^{2}}\left( u \right)}{{{u}^{2}}{{U}_{d,z}}(u)f\left( u \right)}\varphi \left( u \right)=0
\end{equation}
where, without loss of generality, we may take $\psi$ and $\varphi$ to be real,  the prime to denote the derivative with respect to $u$, and ${{U}_{d,z}}\left( u \right)=1-4\gamma d\left( d-1 \right)\xi (u)$. In order to analyse these coupled differential equations, we need to have suitable boundary conditions to be imposed on the conformal boundary $u \to 0$ and on the horizon $u=1$ of the Liftshitz bulk.
The asymptotic behaviors of the scalar  and gauge fields near the boundary $u\to 0$ are:
\begin{eqnarray}\label{E10}
\psi \left( u \right)&\sim & {{\psi}_{1}}{{u}^{{{\Delta }_{-}}}}+{{\psi}_{2}}{{u}^{{{\Delta }_{+}}}} \label{EE3}\\
\varphi (u)&\sim & \left \{ \begin{array}{cc}
   \mu -\rho {{(u/{{r}_{0}})}^{(d-z)}} & 1\le z<~d  \\
   \mu -\rho \ln \left(u{{r}_{0}} \right) & z=d  \\ \end{array} \right \}\label{EE4}
\end{eqnarray}
where, ${{\Delta }_{\pm }}=\frac{\left( z+d \right)\pm \sqrt{{{\left( z+d \right)}^{2}}+4{{m}^{2}}}}{2} $, and ${\psi }_{1}$, ${\psi }_{2}$, $\mu$, and $\rho$ are constant parameters. According to the $\text{AdS/CFT}$ correspondence, $\mu$ will be identified as the chemical potential and $\rho$ as the total charge density in the dual theory. Moreover, $\psi _{1}$ ($\psi _{2}$) can be considered as the source of the dual operator, $O$, with the scaling dimension $\Delta_{-}$ ($\Delta_{+}$). Since
we require the $U(1)$ symmetry to be broken spontaneously, we should turn off the source, i.e., $\psi_{1}=0$. It is obvious that the Breitenlohner-Freedman (BF) bound for the scalar mass in the Lifshitz background becomes ${{m}^{2}}\ge -\frac{1}{4}{{\left( z+d \right)}^{2}}$ for $(d+2)$- dimensions. We take $\Delta=\Delta_{+}$ throughout the paper. At the horizon, $u = 1$, the regularity gives the conditions ${\psi }'(1)=-{{m}^{2}}\psi (1)/(z+d)$ and $\varphi(1)=0$. Furthermore the expansions of the Taylor series  near the horizon are as follows:
\begin{eqnarray}
\psi \left( u \right)=\psi \left( 1 \right)-\psi '\left( 1 \right)\left( 1-u \right)+\frac{1}{2}\psi ''\left( 1 \right){{\left( 1-u \right)}^{2}}\label{EE1}\\
\varphi \left( u \right)=\varphi \left( 1 \right)-\varphi '\left( 1 \right)\left( 1-u \right)+\frac{1}{2}\varphi ''\left( 1 \right){{\left( 1-u \right)}^{2}}\label{EE2}
\end{eqnarray}
From Eqs. \eqref{E9} and \eqref{E8}, and using the regularity conditions $\varphi(1)=0$ and ${\psi }'(1)=-\beta{{m}^{2}}/(z+d)$, we can compute the second derivatives of $\psi(u)$ and  $\varphi(u)$ exactly at the horizon,
\begin{eqnarray}
&{\psi }''\left( 1 \right)=\beta \left[ \frac{{{m}^{2}}}{z+d}\left( 1+\frac{{{m}^{2}}}{2\left( z+d \right)} \right)+\frac{{{\alpha }^{2}}}{2r_{0}^{2z}{{\left( z+d \right)}^{2}}} \right]\\
&{\varphi }''\left( 1 \right)=\alpha \left[ \left( d-z-1 \right)-\frac{{{{{U}'}}_{d,z}}(1)}{{{U}_{d,z}}(1)}-\frac{2{{\beta }^{2}}}{{{U}_{d,z}}(1)(d+z)} \right]
\end{eqnarray}
where, $\alpha =-{\varphi }'\left( 1 \right)<~0$ and $\beta =\psi \left( 1 \right)>~0$.
Thus, we can rewrite Eq. \eqref{EE1} for the scalar field and Eq. \eqref{EE2} for the gauge field as follows:
\begin{eqnarray}
\psi \left( u \right)=\beta +\frac{{{m}^{2}}\beta }{z+d}\left( 1-u \right)+\frac{\beta }{2}\left[ \frac{{{m}^{2}}}{z+d}\left( 1+\frac{{{m}^{2}}}{2\left( z+d \right)} \right)+\frac{{{\alpha }^{2}}}{2r_{0}^{2z}{{\left( z+d \right)}^{2}}} \right]{{\left( 1-u \right)}^{2}}\label{E14}\\
\varphi \left( u \right)=\alpha \left( 1-u \right)-\frac{\alpha }{2}\left[ \left( z-d+1 \right)+\frac{{{{{U}'}}_{d,z}}(1)}{{{U}_{d,z}}(1)}+\frac{2{{\beta }^{2}}}{{{U}_{d,z}}(1)(d+z)} \right]{{\left( 1-u \right)}^{2}}\label{E15}
\end{eqnarray}
We proceed with matching the solutions given by Eqs. \eqref{E14} and \eqref{E15} with Eqs. \eqref{EE3}
and \eqref{EE4} at an intermediate point $u=u_{i}$. Taking into account the following relations:
\begin{eqnarray}
 & {{\left. {{\varphi }^{u\sim 0}} \right|}_{u={{u}_{i}}}}={{\left. {{\varphi }^{u\sim 1}} \right|}_{u={{u}_{i}}}};{{\left. {{{{\varphi }'}}^{u\sim 0}} \right|}_{u={{u}_{i}}}}={{\left. {{{{\varphi }'}}^{u\sim 1}} \right|}_{u={{u}_{i}}}} \\
 & {{\left. {{\psi }^{u\sim 0}} \right|}_{u={{u}_{i}}}}={{\left. {{\psi }^{u\sim 1}} \right|}_{u={{u}_{i}}}};{{\left. {{{{\psi }'}}^{u\sim 0}} \right|}_{u={{u}_{i}}}}={{\left. {{{{\psi }'}}^{u\sim 1}} \right|}_{u={{u}_{i}}}}
\end{eqnarray}
 we obtain
\begin{eqnarray}
\mu -\rho {{\left( \frac{{{u}_{i}}}{{{r}_{0}}} \right)}^{d-z}}&=&\alpha \left( 1-{{u}_{i}} \right)+\frac{{{\left( 1-{{u}_{i}} \right)}^{2}}\alpha }{2}\left[z_{1}+\frac{{{{{U}'}}_{d,z}}(1)}{{{U}_{d,z}}(1)}+\frac{2{{\beta }^{2}}}{{{U}_{d,z}}(1)(d+z)} \right]\label{E18}\\
-(d-z )\rho {{\left( \frac{{{u}_{i}}}{{{r}_{0}}} \right)}^{d-z-1}}&=&\alpha \left[ 1+\left( 1-{{u}_{i}} \right)\left[z_{1}+\frac{{{{{U}'}}_{d,z}}(1)}{{{U}_{d,z}}(1)}+\frac{2{{\beta }^{2}}}{{{U}_{d,z}}(1)(d+z)} \right] \right]\label{E19}\\
\frac{{{\psi }_{2}}{{u}_{i}}^{\Delta }}{\beta}-1&=&\frac{{{m}^{2}}\left( 1-{{u}_{i}} \right)}{z+d} +\frac{{{\left( 1-{{u}_{i}} \right)}^{2}}}{4{{\left( z+d \right)}^{2}}}\left[ 2{{m}^{2}}\left( z+d \right)+{{m}^{4}}+\frac{{{\alpha }^{2}}}{r_{0}^{2z}} \right] \label{E16}\\
\Delta {{\psi }_{2}}{{u}_{i}}^{\Delta -1}&=&-\frac{{{m}^{2}}}{z+d}\beta -\frac{\beta }{2{{\left( z+d \right)}^{2}}}\left[ 2{{m}^{2}}\left( z+d \right)+{{m}^{4}}-\frac{{{\alpha }^{2}}}{r_{0}^{2z}} \right]\left( 1-{{u}_{i}} \right)\label{E17}
\end{eqnarray}
From Eqs. \eqref{E18} and \eqref{E19}, and using Eq. \eqref{q7}, we obtain:
\begin{equation}\label{E21}
\beta^{2}=\kappa \frac{{{U}_{d,z}}(1)(z+d)}{2\left( 1-{{u}_{i}} \right)}{{\left( \frac{{{T}_{c}}}{T} \right)}^{\frac{d}{z}}}\left[ 1-{{\left( \frac{T}{{{T}_{c}}} \right)}^{\frac{z}{d}}} \right]
\end{equation}
where, $\kappa $ and $T_{c}$ are given by:
\begin{eqnarray}
&&\kappa = 1+\left( 1-{{u}_{i}} \right)\left( \left( z-d+1 \right)+\frac{U{{\text{ }\!\!'\!\!\text{ }}_{d,z}}(1)}{{{U}_{d,z}}(1)} \right)\\
&&{{T}_{c}}=\frac{\left( z+d \right)}{4\pi }{{\left[ \frac{\left( d-z \right)u_{i}^{d-z-1}\rho }{\kappa \tilde{\alpha }} \right]}^{\frac{z}{d}}}
\end{eqnarray}
For $T\sim {{T}_{c}}$, Eq. \eqref{E21} leads to
\begin{equation}
\beta =\sqrt{\kappa \frac{{{U}_{d,z}}(1)(z+d)}{2\left( 1-{{u}_{i}} \right)}\left[ 1-{{\left( \frac{T}{{{T}_{c}}} \right)}^{\frac{z}{d}}} \right]}
\end{equation}
On the other hand, by using Eqs. \eqref{E18} and \eqref{E19}, we have:
\begin{eqnarray}\label{E20}
{{\psi }_{2}}&=&\beta \frac{~{{u}_{i}}^{1-\Delta }\left[ {{m}^{2}}\left( {{u}_{i}}-1 \right)-2\left( z+d \right) \right]}{\left( z+d \right)\left[ \left( \Delta -2 \right){{u}_{i}}-\Delta  \right]}\\
 \nonumber \tilde{\alpha }&=&~\frac{\alpha }{r_{0}^{z}}={{\left[ m^{2}\left( {{m}^{2}}+2\left( z+d \right)\left( \frac{2-{{u}_{i}}}{1-{{u}_{i}}} \right) \right)+\frac{2\Delta \delta \left( z+d \right)}{\left( 1-{{u}_{i}} \right)}\right]}^{\frac{1}{2}}}
\end{eqnarray}
where, $\delta =\left( {{m}^{2}}\left( {{u}_{i}}-1 \right)-2\left( z+d \right) \right)/\left( \left( \Delta -2 \right){{u}_{i}}-\Delta  \right)$.  Since the
 (critical) temperature must be positive, we need to consider the following constraint:
\begin{equation}
Max\left\{ 0,{{u}_{m}},{{u}_{\gamma }} \right\}<{{u}_{i}}<1
\end{equation}
where,
\begin{eqnarray}
{{u}_{\gamma }}&=&\frac{8\left( \left( z+1 \right)d-{{z}^{2}}+2-z \right)\left( d+z \right)\left( d-1 \right)\gamma +\left( d+1 \right)\left( -2-z+d \right)}{8\left( d+z \right)\left( \left( 1/2+z \right)d+1-{{z}^{2}} \right)\left( d-1 \right)\gamma +\left( d+1 \right)\left( -z+d-1 \right)}\\
\nonumber {{u}_{m}}&=&\frac{\Delta \left[ {{m}^{4}}+6{{m}^{2}}\left( z+d \right)+4{{\left( z+d \right)}^{2}} \right]}{{{m}^{2}}\left( \Delta -1 \right)\left( {{m}^{2}}+4z+4d \right)-m\sqrt{\left[ {{m}^{2}}{{\left( {{m}^{2}}+4z+4d \right)}^{2}}-8\Delta \left( \Delta -2 \right){{\left( z+d \right)}^{3}} \right]}}
\end{eqnarray}
Therefore, we cannot choose an arbitrary value for $u_{i}$ because the matching point $u_{i}$ depends on the Lifshitz
scaling $z$, dimension $d$,  scalar mass $m$, and Weyl coupling $\gamma$. These constraints can also be used to ensure that $\beta$ is real. It is interesting that when $\gamma=0$ in $(d+2)$-dimensions, the corresponding results recover the ones in Ref. ~\cite{ref5}. In addition, when we choose  $\gamma\ne0$, $d=3$, and $u_{i}=0.5$, the values for critical temperature approximate those in Ref. ~\cite{ref6} for $z=1$. The values for  critical temperature are computed below for various selected values  of the Lifshitz
scaling, $z$. The plots of temperature versus Lifshitz scaling for different values of the Weyl coupling are presented in Figure ~\ref{pic1}.
 These plots show that the value for critical temperature, $T_{c}$, decreases as the Lifshitz scaling, $z$, increases but it decreases as the Weyl coupling decreases when $0<z \le 1$ and $-0.08<\gamma\le0$. Furthermore, in the case $z-2<d<2z-2$,  the values for critical temperature, $T_{c}$, are reported in Table ~\ref{table1} for  condensations with different values of Weyl coupling, $\gamma$, and dynamical exponent, $z$,  when  $d=10$. Clearly,  the gradual increase in the Weyl coupling  helps an easier  condensation to occur.
\begin{table}[h]\label{table1}
\begin{center}
 \begin{tabular}{|l |c |r|r|r|r|r|}
\hline
  & $\gamma $ &$0.04$ & $0.05$ & $0.06$ & $0.07$ & $0.08$  \\[-0.25ex]
       \raisebox{1.5ex}{$ z=7 $}  &${{T}_{c}}$ & $0.20364{{\rho }^{\frac{7}{10}}}$ & $0.26784{{\rho }^{\frac{7}{10}}}$ & $0.35834{{\rho }^{\frac{7}{10}}}$ &$0.50364{{\rho }^{\frac{7}{10}}}$ & $0.80202{{\rho }^{\frac{7}{10}}}$ \\
          \hline
            & $\gamma $ &$0.029$ & $0.032$ & $0.035$ & $0.038$ & $0.041$  \\[-0.25ex]
       \raisebox{1.5ex}{$ z=8 $}  &${{T}_{c}}$ & $0.31504{{\rho }^{\frac{4}{5}}}$ & $0.38425{{\rho }^{\frac{4}{5}}}$ & $0.47793{{\rho }^{\frac{4}{5}}}$&$0.61410{{\rho }^{\frac{4}{5}}}$ & $0.83527{{\rho }^{\frac{4}{5}}}$ \\
          \hline
          & $\gamma $ &$0.022$ & $0.023$ & $0.024$ & $0.025$ & $0.026$  \\[-0.25ex]
       \raisebox{1.5ex}{$ z=9 $}  &${{T}_{c}}$ & $0.39497{{\rho }^{\frac{9}{10}}}$ & $0.45733{{\rho }^{\frac{9}{10}}}$ & $0.53710{{\rho }^{\frac{9}{10}}}$&$0.64313{{\rho }^{\frac{9}{10}}}$ & $0.79175{{\rho }^{\frac{9}{10}}}$ \\
          \hline
          \end{tabular}
\end{center}
    \caption{\label{table1}Values of critical temperature $T_{c}$ for different values of the Weyl coupling,  $\gamma$, and Lifshitz scaling, $z$.}
\end{table}
We can also write the expression for the condensation operator $\left\langle O \right\rangle =~{{\psi }_{2}}{{r}_{H}}^{\Delta }$ near the critical temperature $T\sim {{T}_{c}}$ as in the following:
\begin{equation}
 {{\left\langle O \right\rangle }^{\frac{1}{\Delta }}}~=\lambda {{\left( \frac{4\pi {{\left[ {{U}_{d,z}}\left( 1 \right) \right]}^{\frac{z}{2\Delta }}}}{d+z} \right)}^{\frac{1}{z}}}{{T}_{c}}^{\frac{1}{z}}{{\left[ 1-{{\left( \frac{T}{{{T}_{c}}} \right)}^{\frac{z}{d}}} \right]}^{\frac{1}{2\Delta }}}
 \end{equation}
with
\begin{equation}
\lambda ={{\left[ \frac{\delta \sqrt{\kappa }(z+d)~{{u}_{i}}^{1-\Delta }}{\sqrt{2\left( 1-{{u}_{i}} \right)}} \right]}^{\frac{1}{\Delta }}}
\end{equation}
\begin{figure}[tbp]
\includegraphics[scale=.4]{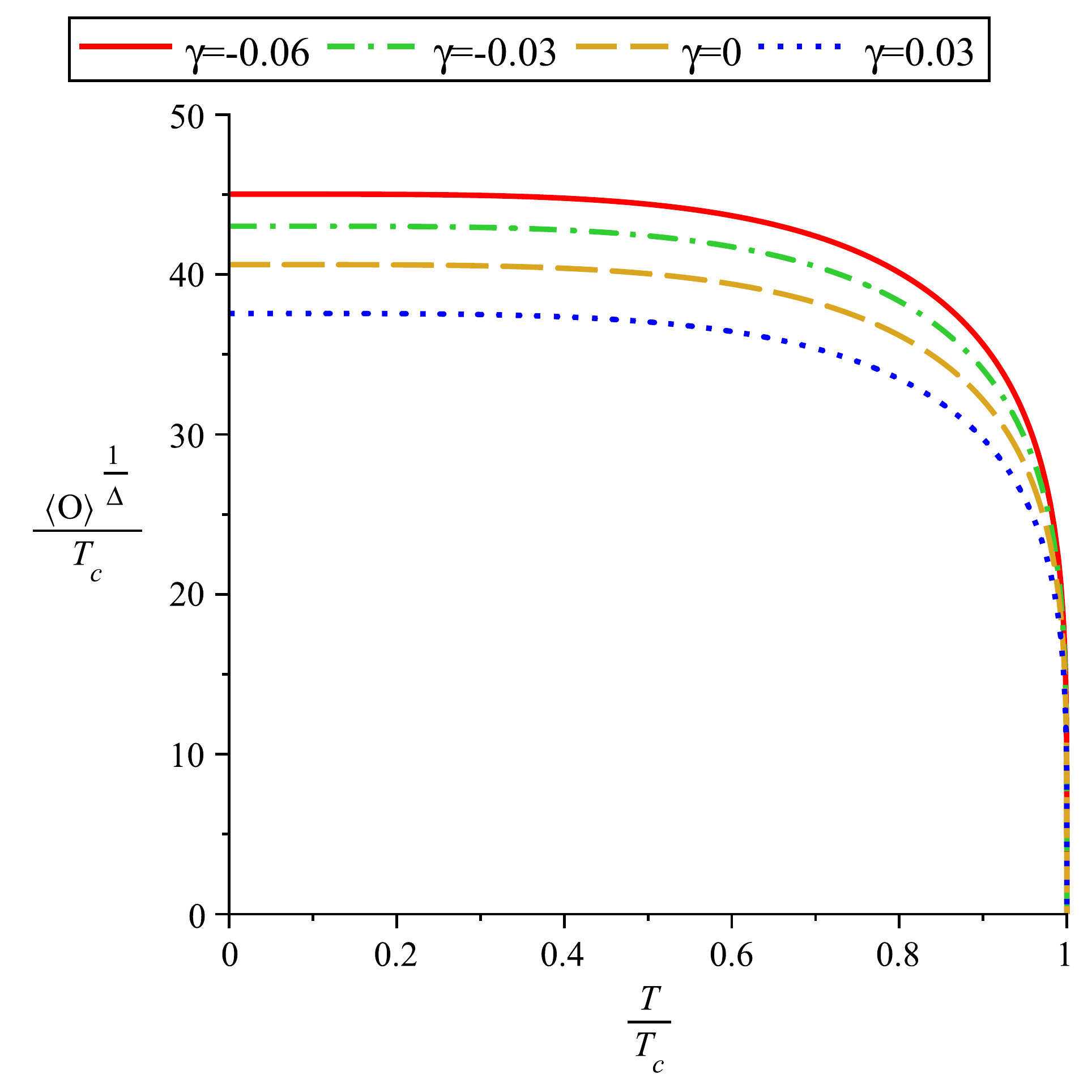}
\includegraphics[scale=.4]{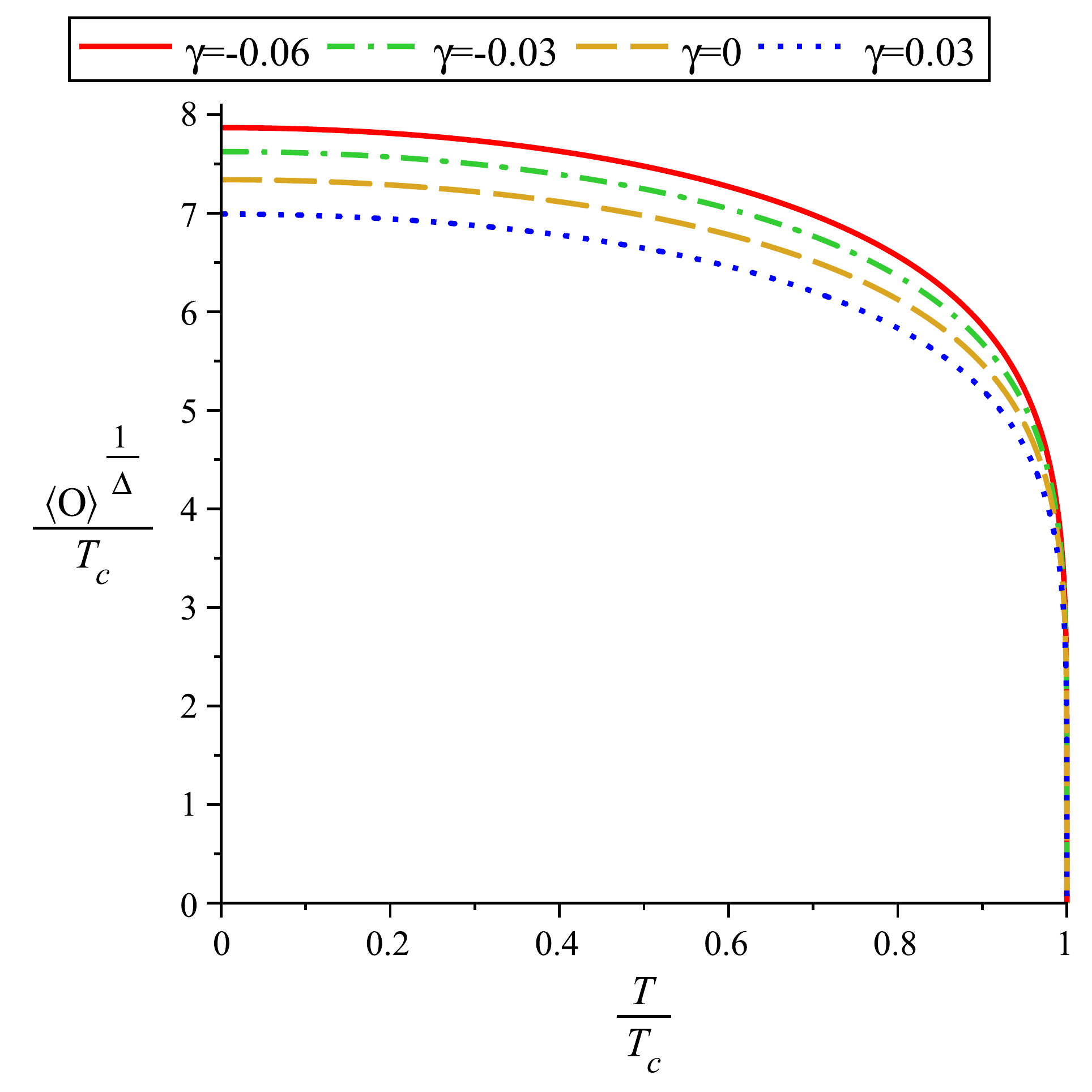}
\caption{\label{fig2}Value of  condensate as a function of  temperature for the
solutions with $z=0.5$ (left) and $z=1$ (right). In both plots, $u$ and $m^2$ are chosen to be (0.5) and (-0.5), respectively.  }
\end{figure}
Fig. (\ref{fig2}) shows the dependence of  condensation on the coupling to the Weyl correction for different values  of $z$. Comparison of  both sides of  Fig. ~\ref{fig2} reveals  that the gap becomes smaller as the Liftshitz scaling, $z$, and the Weyl coupling, $\gamma$, increase.

\section{Effect of the external magnetic field on Weyl superconductors}\label{cap3}
\begin{figure}[tbp]
\centering
\includegraphics[scale=.5]{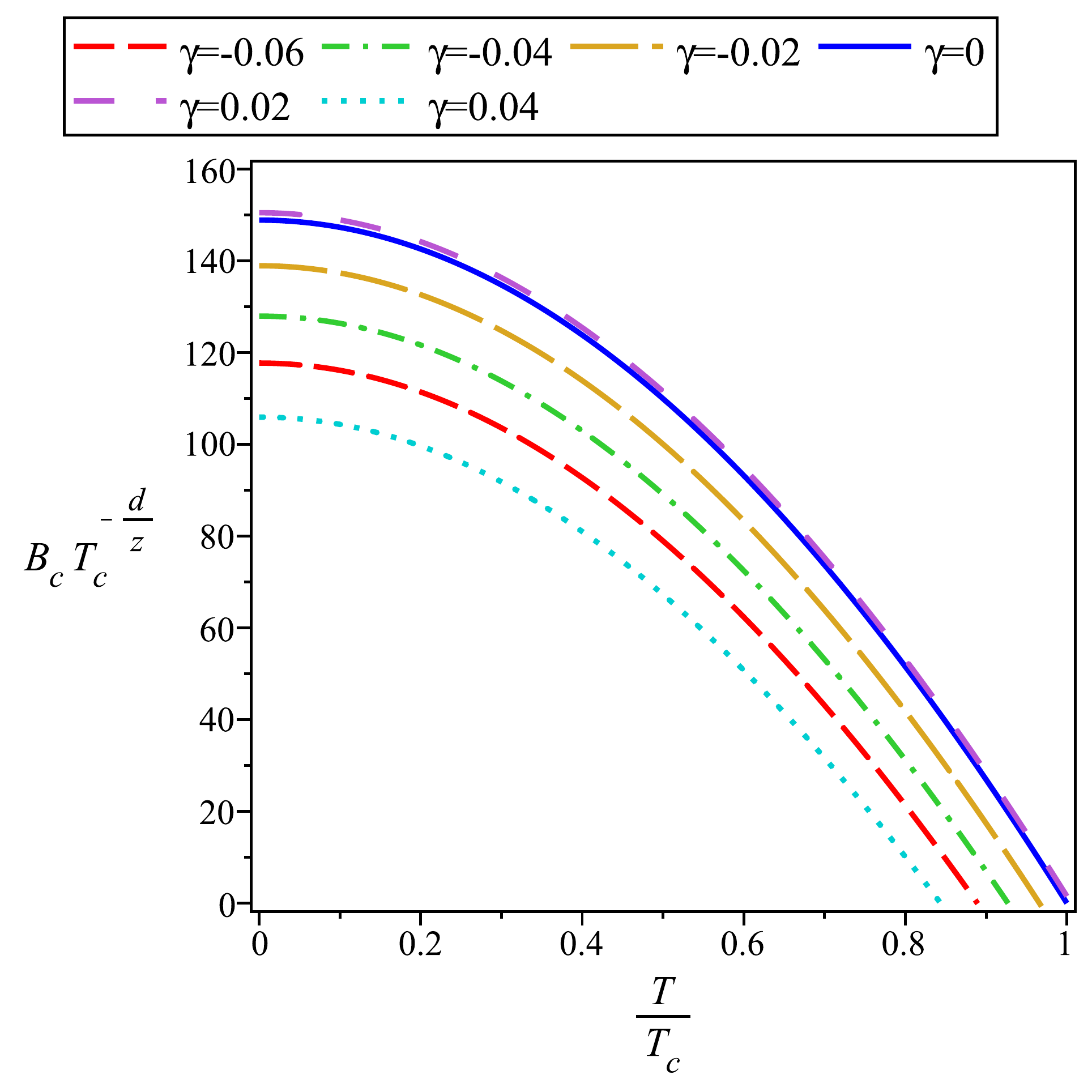}
\caption{Critical magnetic field as a function of temperature for different values of Weyl couplings for $z=1$.
We assume $m^2 = -0.5$,  $d = 2$, and $u_{i} = 1/2$. }\label{fig5}
\end{figure}
\begin{figure}[tbp]
\includegraphics[scale=.4]{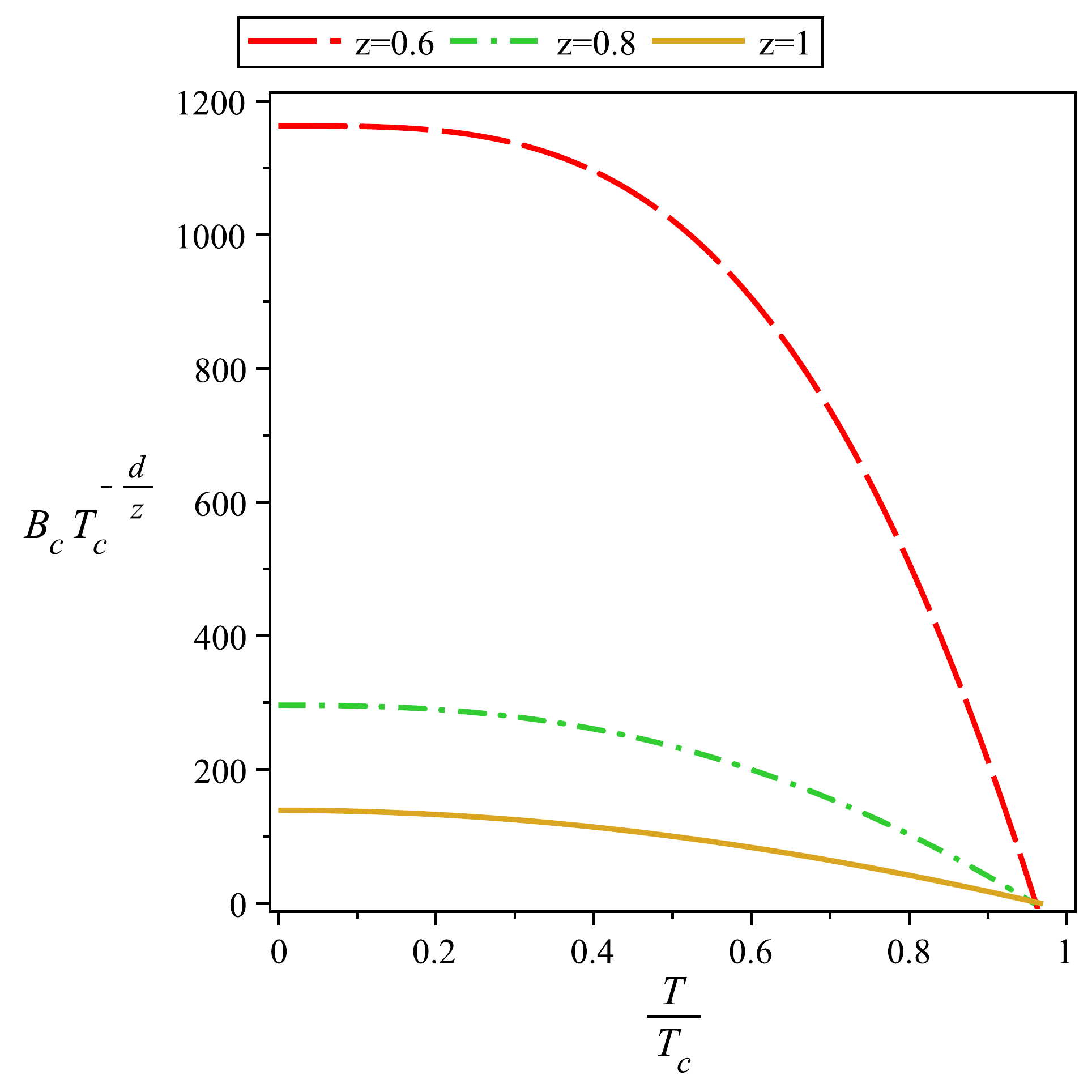}
\includegraphics[scale=.4]{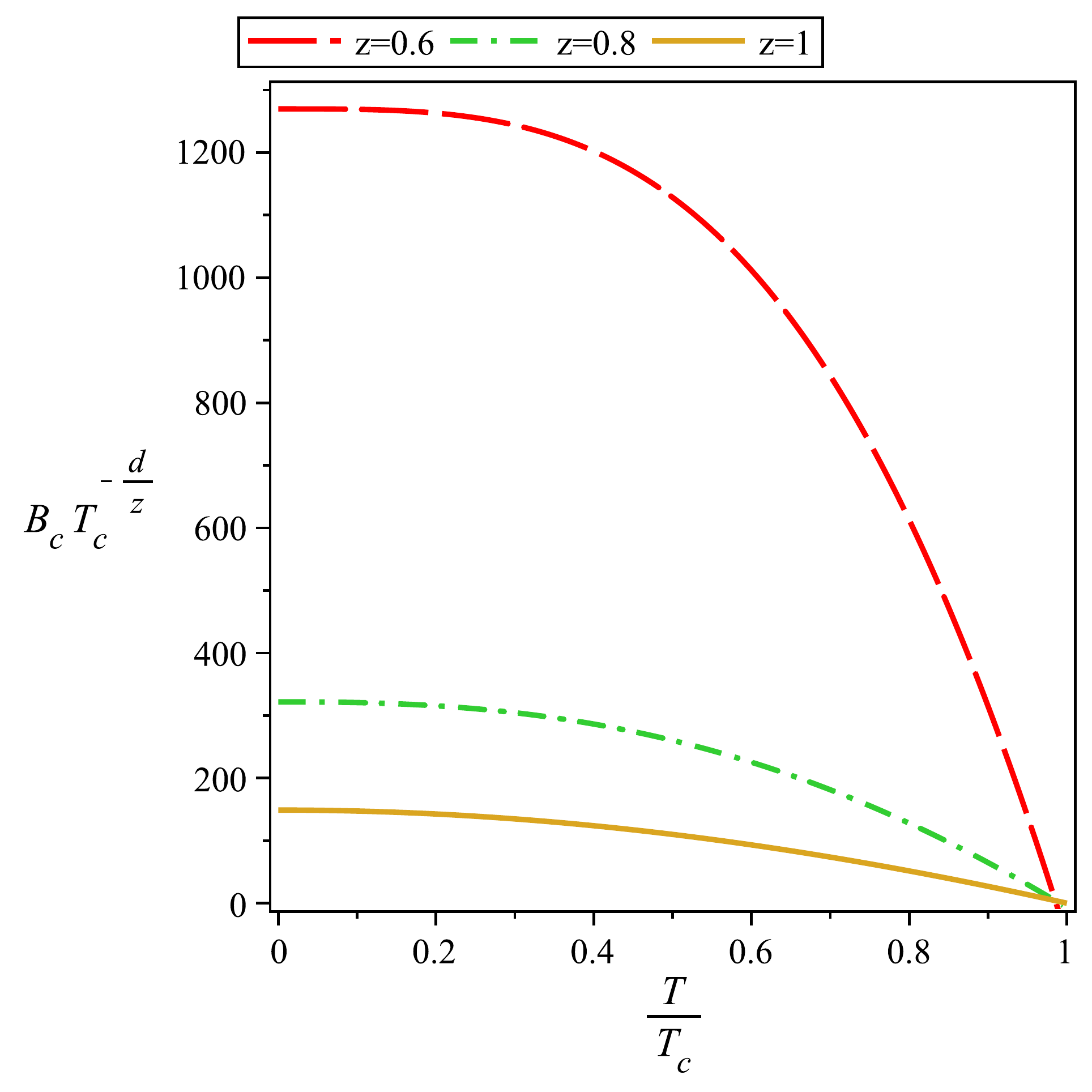}
\includegraphics[scale=.4]{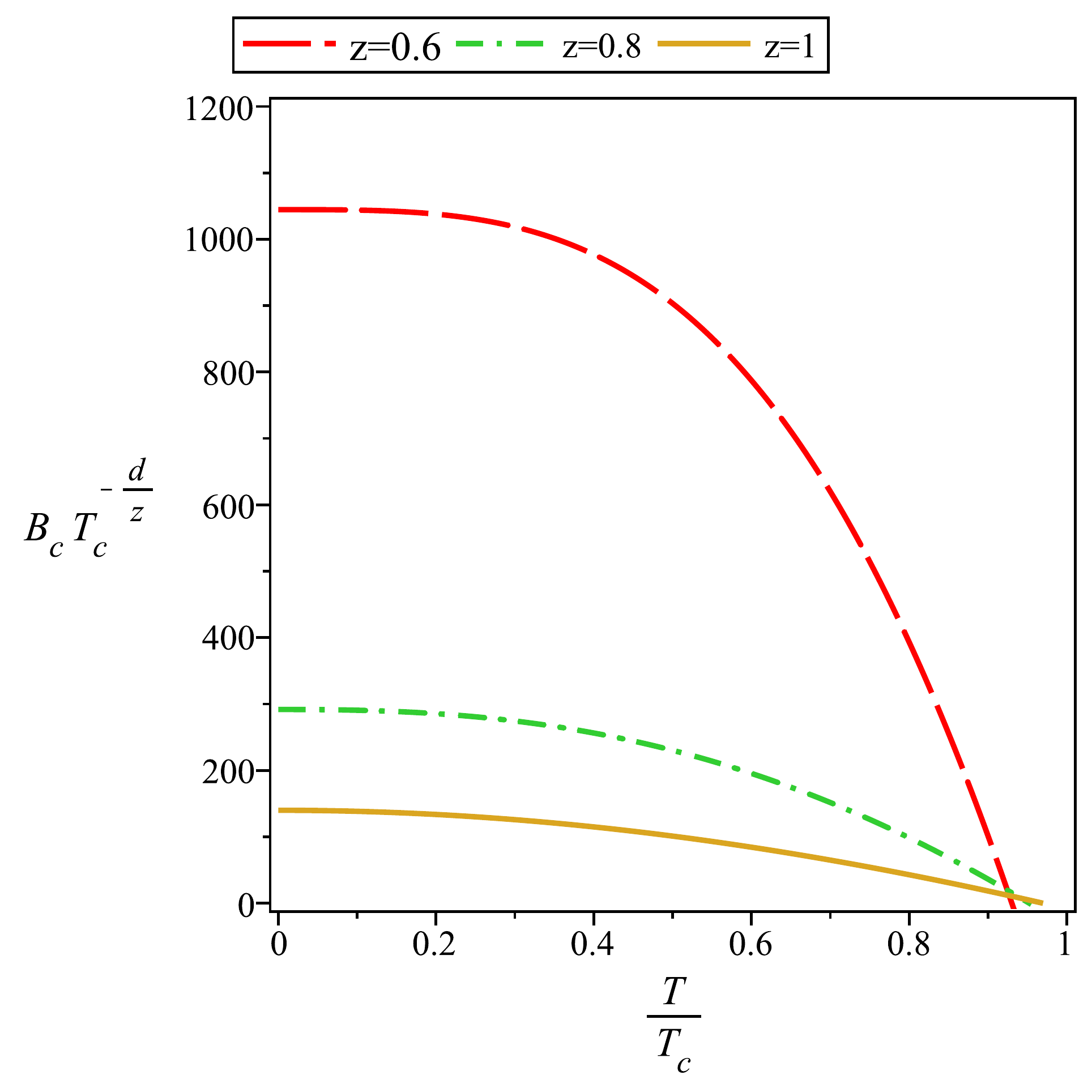}
\includegraphics[scale=.4]{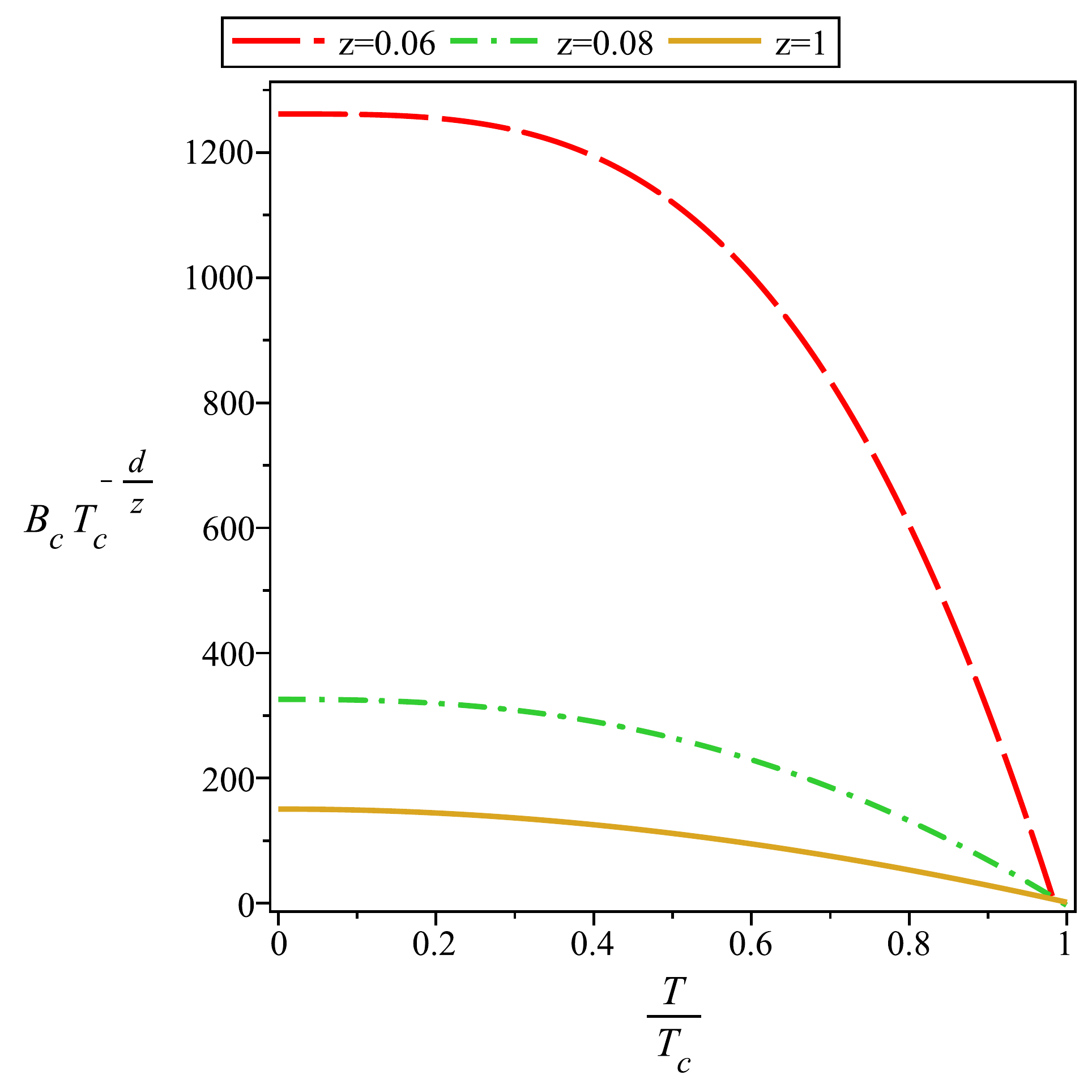}
\caption{The value of the critical magnetic field as a function of the temperature for the
solutions with $\gamma={-0.02, 0, 0.02, 0.03}$. The $u$, $d$ and $m^2$ are chosen to be (0.5), (2), and (-0.5), respectively.}\label{fig6}
\end{figure}
This Section  investigates the effects of an external static magnetic field, $B$. For this purpose,  a magnetic
field may be placed with  other fields in the bulk. From the AdS/CFT correspondence, it follows that the asymptotic value of this magnetic field corresponds to a magnetic
field added to the boundary field theory. Therefore, we make the following ansats  ~\cite{D. Roychowdhury,D. Roychowdhury1,T. Albash,T. Albash1}:
\begin{eqnarray}\label{form-1}
{{A}_{\mu }}d{{x}^{\mu }}&=&\varphi (u)dt+\left( By \right)dx\nonumber\\
 \psi &=& \psi(y,u)
\end{eqnarray}
This leads to the following equation of motion for the scalar field $\psi(y,u)$:
\begin{eqnarray}\label{form-2}
&& {{\psi }^{''}}\left( u,y \right)+{{\psi }^{'}}\left( u,y \right)\left[ \frac{{{f}^{'}}\left( u \right)}{f\left( u \right)}-\frac{d+z-1}{u} \right]+\psi \left( u \right)\left[ \frac{{{u}^{2z-2}}{{\varphi }^{2}}\left( u \right)}{r_{0}^{2z}{{f}^{2}}\left( u \right)}-\frac{{{m}^{2}}}{{{u}^{2}}f\left( u \right)} \right] \\
\nonumber &&+ \frac{1}{r_{0}^{2}f(u)^2}\left[ \partial_{y}^2-B^{2}y^{2}\right]\psi(y,u) =0
\end{eqnarray}
One can solve this equation by taking the following separable form for the scalar field:
\begin{equation}\label{form-3}
\psi(y,u)=Q(y)P(u)
\end{equation}
Substituting Eq. \eqref{form-3} into Eq. \eqref{form-2} yields:
\begin{eqnarray}\label{form-4}
&&\frac{P''(u)}{P(u)}+\frac{P'(u)}{P(u)}\left[ \frac{f'(u)}{f(u)}+\frac{1-d-z}{u}\right] +\left[ \frac{u^{2z-2}\varphi(u)^{2}}{r_{0}^{2z}f(u)}-\frac{m^2}{u^{2}f(u)}\right] \\
\nonumber &&-\frac{1}{r_{0}^{2}f(u)}\Big[-\frac{Q''(y)}{Q(y)}+B^{2}y^{2}\Big] =0
\end{eqnarray}
The $y$ dependent part of Eq. \eqref{form-4} yields the quantum harmonic oscillator in one dimension with the relevant frequency determined by $B$ as follows:
\begin{equation}\label{form-5}
Q''(y)+B^{2}y^{2}Q(y)=c_{n}BQ(y)
\end{equation}
where, $c_{n}=2n+1$ is a constant. In the stable state and in the lowest mode ($n=0$), the $u$ dependent part of Eq. \eqref{form-4} can be expressed by:
\begin{eqnarray}\label{form-6}
P''(u)+P'(u)\left[ \frac{f'(u)}{f(u)}+\frac{1-d-z}{u}\right] +P(u)\Big[  \frac{u^{2z-2}\varphi(u)^{2}}{r_{0}^{2z}f(u)}
-\frac{m^2}{u^{2}f(u)}-\frac{B}{r_{0}^{2}f(u)^2}\Big]  =0
\end{eqnarray}
 Using the regularity condition, $\varphi(1)=0$, one can obtain
the following relation at the horizon ($u=1$):
\begin{equation}\label{form-7}
P'(1)=(\frac{-m^2}{z+d}+\frac{B}{r_{0}^{2}})P(1)
\end{equation}
Moreover, based on Eq. \eqref{form-6}, Eq. \eqref{form-7}, and the regularity condition for $\varphi$, we have:
\begin{eqnarray}\label{form-8}
P''(1)&=&\frac{1}{(z+d)^2}\Big[ m^{2}(z+d)+\frac{m^4}{2}+\frac{\varphi'(1)^{2}}{2r_{0}^{2z}}
+\frac{Bm^{2}}{r_{0}^2}+\frac{B^2}{2r_{0}^4}\Big] P(1)
\end{eqnarray}
On the other hand,  at the boundary ($u\longrightarrow0$), the asymptotic solution of Eq. \eqref{form-6} can be written as:
\begin{equation}\label{form-9}
P(u)=I_{-}u^{\Delta_{-}}+I_{+}u^{\Delta_{+}}
\end{equation}
where, as previously  chosen, we set $I_{-} = 0$  and $\Delta=\Delta_{+}$. In order to find the value of the critical magnetic field, we need to take the matching method of solutions near the horizon and the boundary. To do this, we consider the expansion of the $P(u)$ near the horizon ($u=1$) as follows:
\begin{equation}\label{form-9}
P(u)=P(1)+P'(1)(1-u)+\frac{1}{2}P''(1)(1-u)^2+...
\end{equation}
 Puting  Eqs. (\ref{form-7}) and (\ref{form-8}) into Eq. (\ref{form-9}), we have:
\begin{eqnarray}\label{form-10}
&& \nonumber P(u)=P(1)+(\frac{-m^2}{z+d}+\frac{B}{r_{0}^2})P(1)(1-u)+\frac{1}{(z+d)^2}\Big[  m^{2}(z+d)+\frac{m^{4}}{2}+\\
&& \frac{\varphi'(1)^{2}}{2r_{0}^{2z}}+\frac{Bm^{2}}{r_{0}^2}+\frac{B^2}{2r_{0}^4}\Big] P(1)(1-u)^2
\end{eqnarray}
Matching this solution with Eq. \eqref{form-9} with $I_{-}= 0$ at some intermediate point $u = u_{i}$, we get the following relations:
\begin{eqnarray}\label{form-11}
 Iu_{i}^{\Delta }\text{ }&=&P(1)\left[ 1+(\frac{-{{m}^{2}}}{z+d}+\frac{{{B}^{2}}}{r_{0}^{2}})(1-{{u}_{i}}) \right]+
 \frac{P\left( 1 \right){{\left( 1-{{u}_{i}} \right)}^{2}}}{2{{(z+d)}^{2}}}\Big[ {{m}^{2}}(z+d)+\frac{{{m}^{4}}}{2}\\
\nonumber && + \frac{{\varphi }'{{(1)}^{2}}}{2r_{0}^{2z}}+
  \frac{B{{m}^{2}}}{r_{0}^{2}}+\frac{{{B}^{2}}}{2r_{0}^{4}} \Big]\\
 I\Delta u_{i}^{\Delta -1}&=&-P(1){{u}_{i}}\left[ \frac{-{{m}^{2}}}{z+d}+\frac{B}{r_{0}^{2}} \right]-\frac{P(1)(1-{{u}_{i}})}{{{(z+d)}^{2}}}\big[ {{m}^{2}}(z+d)+\frac{{{m}^{4}}}{2}+\frac{{\varphi }'{{(1)}^{2}}}{2r_{0}^{2z}}\\
 \nonumber &&+\frac{B{{m}^{2}}}{r_{0}^{2}}+\frac{{{B}^{2}}}{2r_{0}^{4}}\big]
\end{eqnarray}
The above set of equations yields the following solution to the magnetic field $B$.
\begin{equation}\label{form-13}
B=\frac{r_{0}^{2}}{\zeta}\left[ \sqrt{\omega+\zeta^{2}(\frac{\varphi'(1)}{r_{0}^{z}})^{2}}-\delta\right]
\end{equation}
with
\begin{eqnarray}\label{form-14}
\zeta &=& \left[ (\Delta -2)u_{i}-\Delta\right] (u_{i}-1)\nonumber\\
\omega &=&2(z+d)\left[  2u_{i}^{2}(z+d)-(m\zeta)^{2}\right] \nonumber\\
\delta &=&\left[ (z+d)-m^{2}(u_{i}-1)\right] (2u_{i}+\Delta(1-u_{i})) \\
\nonumber &+&\Delta (1-u_{i})(z+d)
\end{eqnarray}
The value of the magnetic field is assumed to be very close to the critical magnetic field strength, $B_c$. Since the condensate is so small in this situation, one can ignore the quadratic terms in $\psi$ so that  Eq. \eqref{E8} reduces to:
 \begin{equation}\label{form-15}
\varphi''(u)+\varphi'(u)\left[ \frac{(z-d+1)}{u}+\frac{U'_{d,z}(u)}{U_{d,z}(u)}\right] =0
\end{equation}
 Integrating the above equation in the interval $[1, u]$, and using the asymptotic boundary condition
for $\varphi$ in Eq. \eqref{form-9}, one can obtain:
\begin{equation}\label{form-16}
\varphi'(u)=\frac{C (u^{-z+d-1})}{4d\gamma (d-1)(d+2-z)u^{d+z+4}-8\gamma z (-1+z)(d-1)u^{4}-d-1}
\end{equation}
where,
\begin{equation}
 C =\frac{(d+1)(d-z) \rho}{r_{0}^{d-z}}
\end{equation}
Therefore, we compute $\varphi'(u)$ at the horizon ($u=1$) as follows:
 \begin{equation}\label{form-22}
 \varphi' (1)=\frac{1}{r_{0}^{d-z}}\left[ \frac{(d+1)(d-z)\rho}{4(d-1)(d+1)(d-2z+2)\gamma -d-1}\right]
 \end{equation}
Consequently, Eqs. \eqref{form-13} and \eqref{form-22} give us the following value for the critical magnetic field.
 \begin{equation}\label{form-23}
 B_{c}=\frac{(\frac{4\pi T}{d+z})^{\frac{2}{z}}(\frac{T_c}{T})^{\frac{d}{z}}}{\zeta}\left[ \sqrt{\omega (\frac{T}{T_c})^{\frac{2d}{z}}+(\frac{\zeta \eta}{(T_c)^{\frac{d}{z}}})^{2}}-\delta(\frac{T}{T_c})^{\frac{d}{z}}\right]
 \end{equation}
 with
 \begin{equation}\label{form-24}
 \eta= (\frac{d+z}{4\pi})^{\frac{d}{z}}\left[ \frac{(d+1)(d-z)\rho}{4(d-1)(d+z)(d-2z+2)\gamma -d-1}\right]
 \end{equation}
This result reveals the dependence of  critical magnetic field on  Weyl coupling, $\gamma$. However, we find that the critical magnetic field, $B_{c}$ decreases as $T/T_{c}$ rises. Also, from Fig. ~\ref{fig5} we find that the critical magnetic field vanishes at $T<T_{c}$  for $\gamma<0$, at $T=T_{c}$ for $\gamma=0$ and for $\gamma>0$ at $T<T_{c}$. Moreover, Fig. ~\ref{fig6} shows that the critical magnetic field, $B_{c}$, decreases as we amplify $z$ for the fixed values of  $-0.02 \le \gamma \le 0.03$. This implies that the dynamical exponent, $ z$,  affects the critical magnetic field.

\section{\label{sec:level2}Electrical conductivity  }\label{cap5} 
\begin{figure}[tbp] 
\centering
\includegraphics[scale=.7]{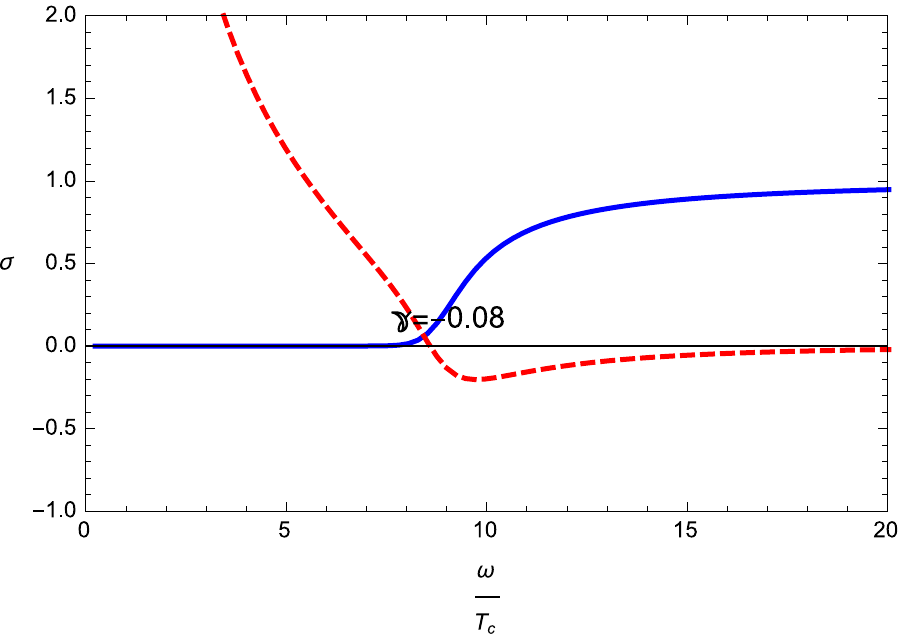}
\includegraphics[scale=.7]{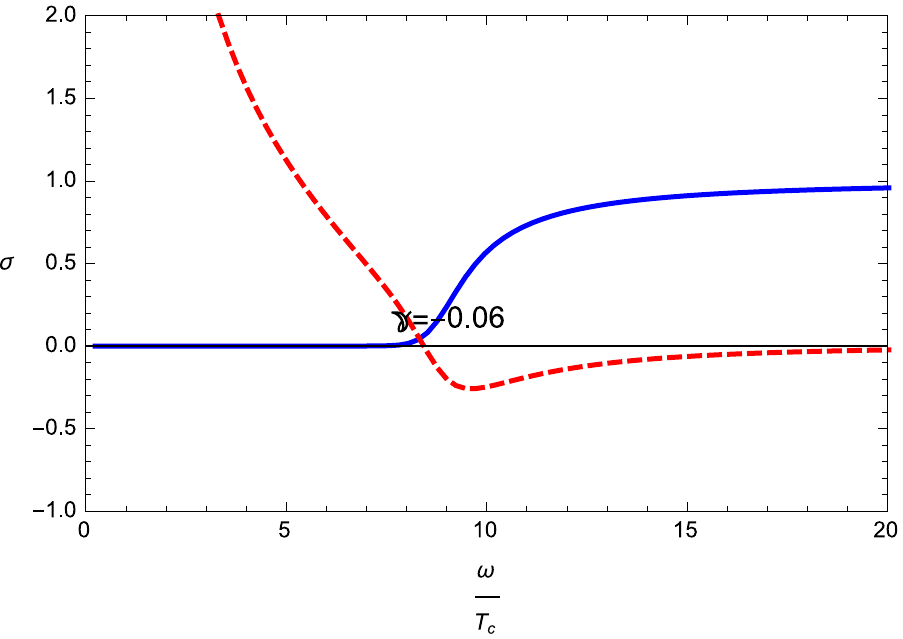}
\includegraphics[scale=.7]{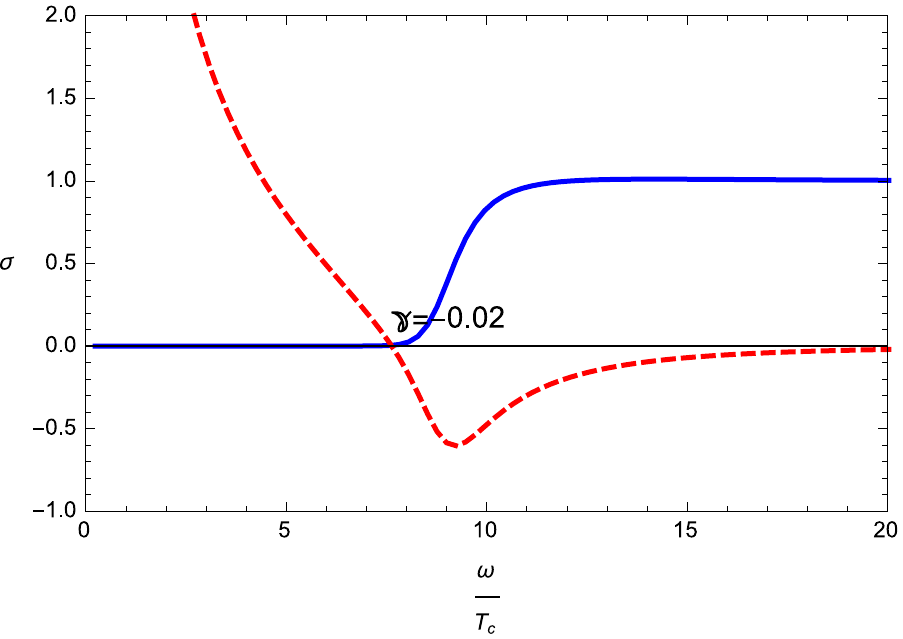}
\includegraphics[scale=.7]{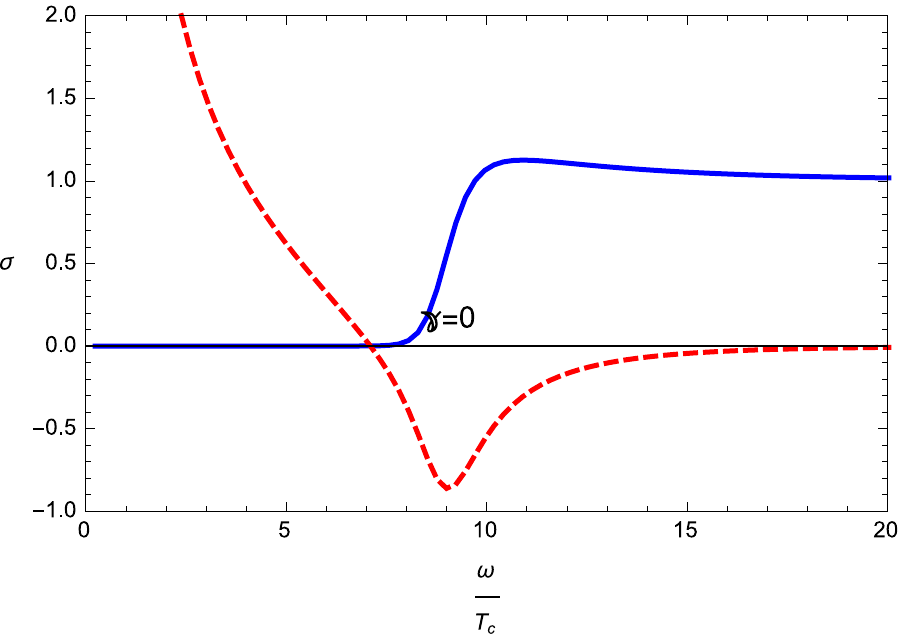}
\includegraphics[scale=.7]{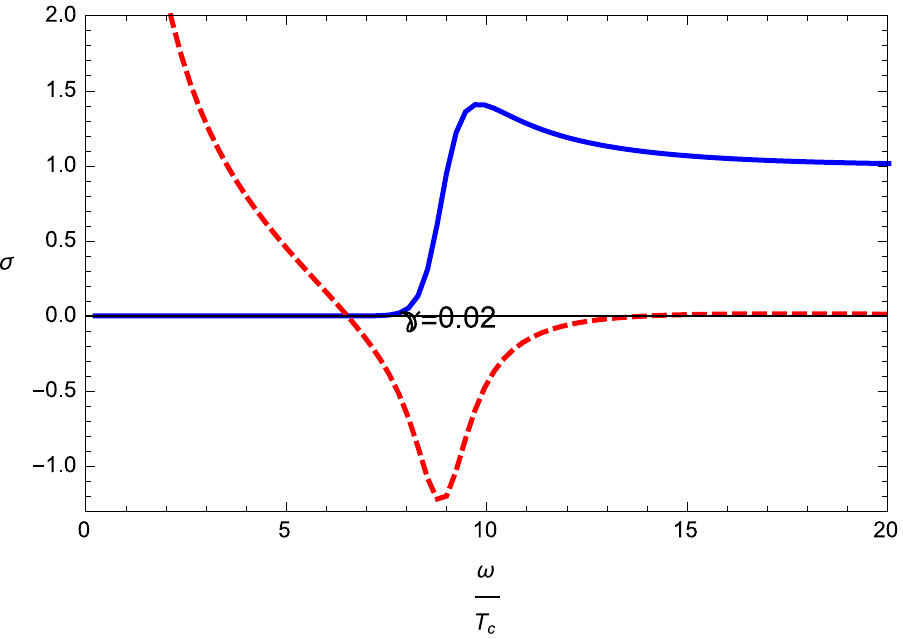}
\includegraphics[scale=.7]{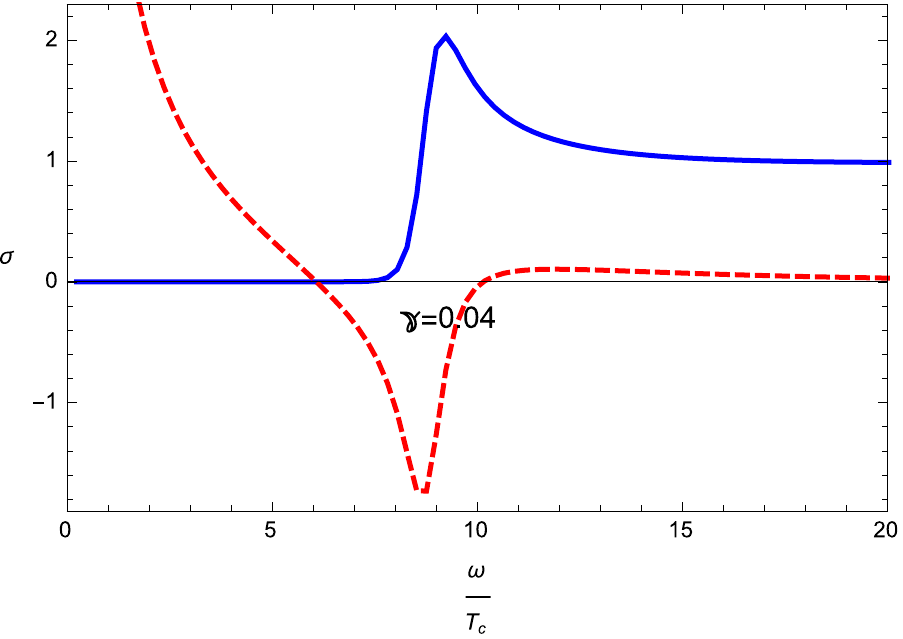}
\includegraphics[scale=.7]{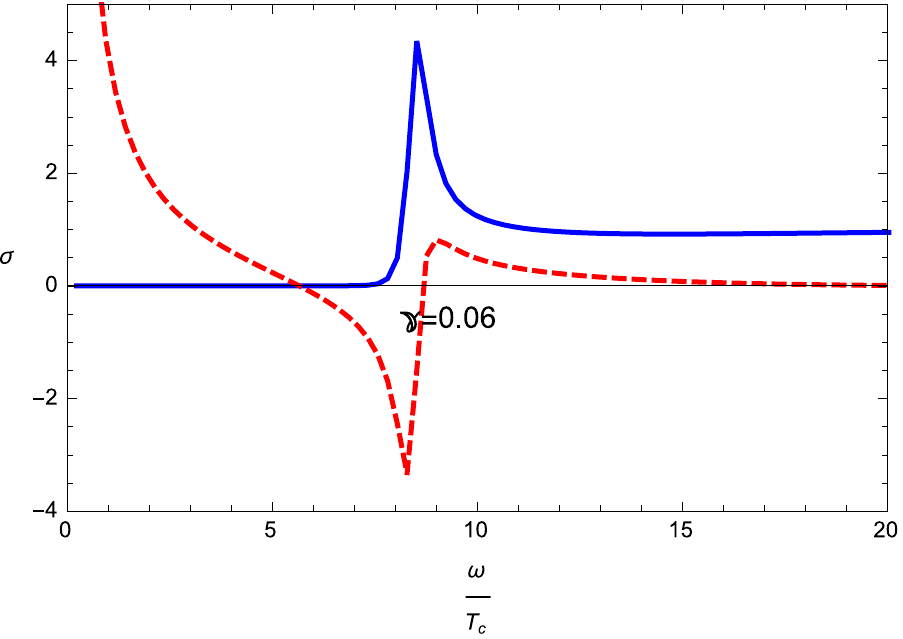}
\includegraphics[scale=.7]{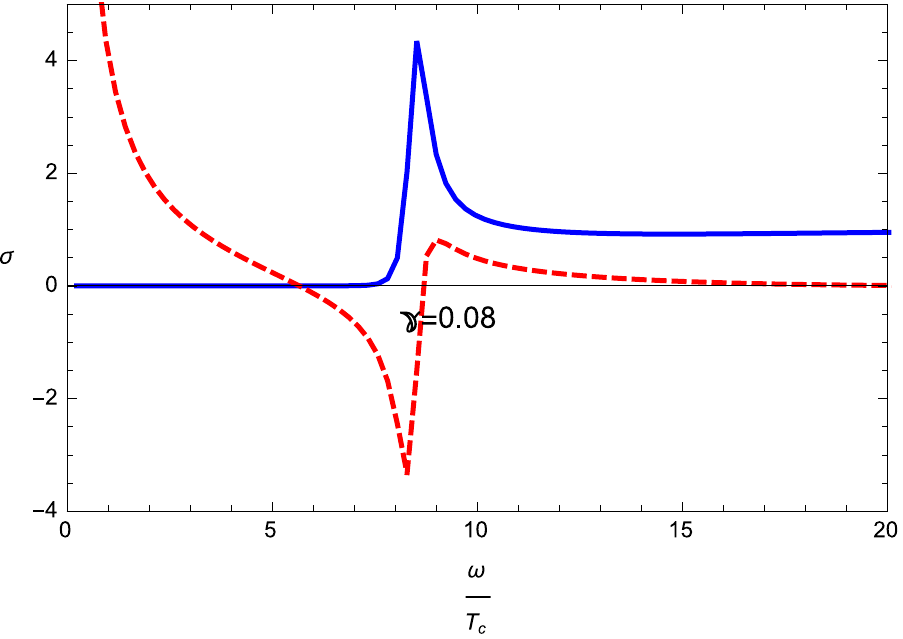}
\caption{The real (solid blue curve) and imaginary (dashed red curve) part of the AC conductivity versus frequency of the Weyl
model at $T/Tc \approx 0.105151$ with $\Delta=2$, $z=1$, and $d=2$ for different $\gamma=-0.08,-0.06,-0.02,0,0.02,0.04,0.06,0.08$ from top to bottom.}\label{fig7}
\end{figure}

\begin{figure}[tbp]
\centering
\includegraphics[scale=.7]{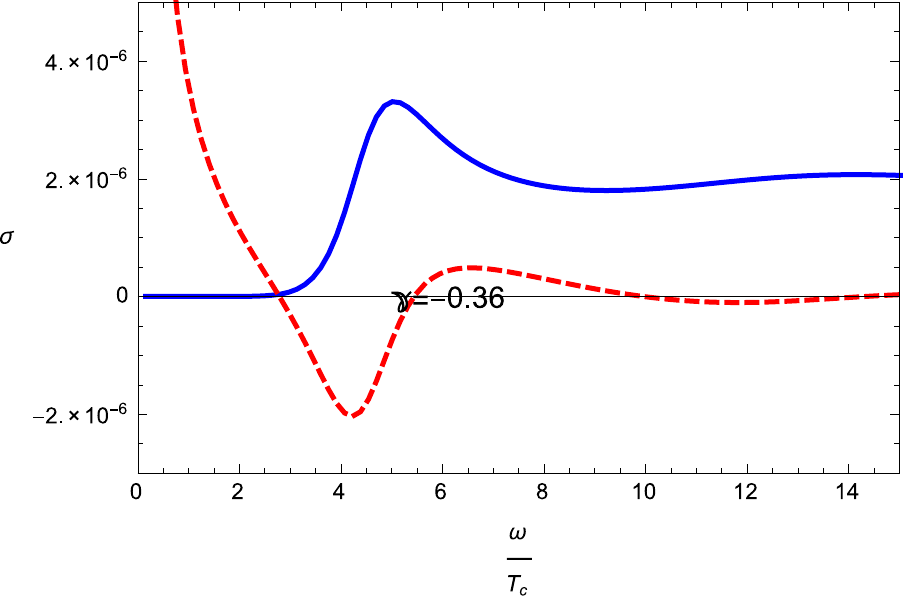}
\includegraphics[scale=.7]{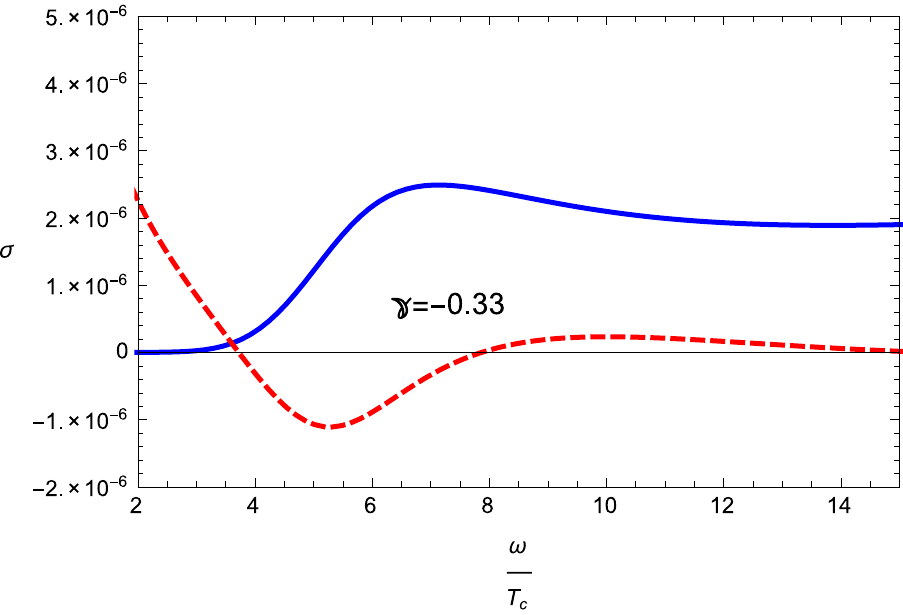}
\includegraphics[scale=.7]{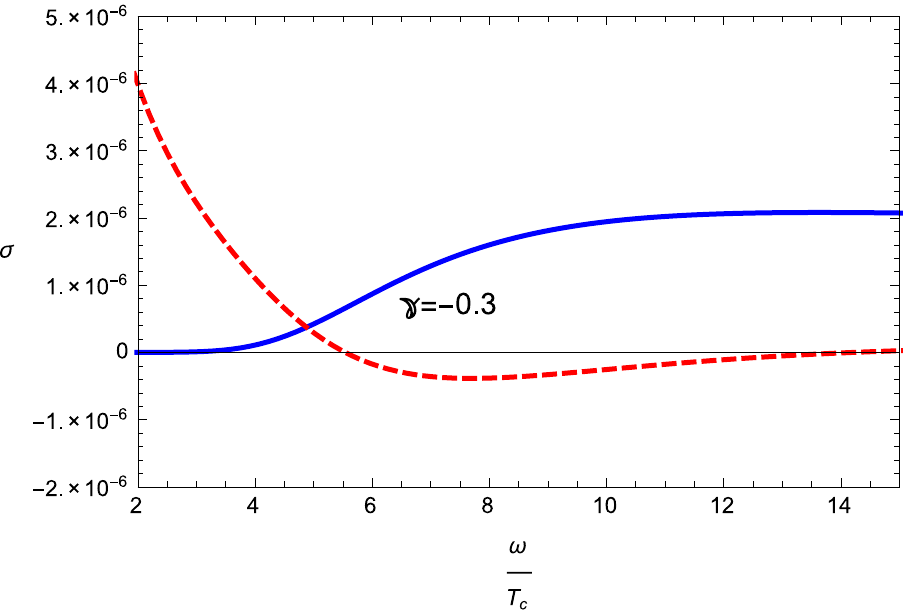}
\includegraphics[scale=.7]{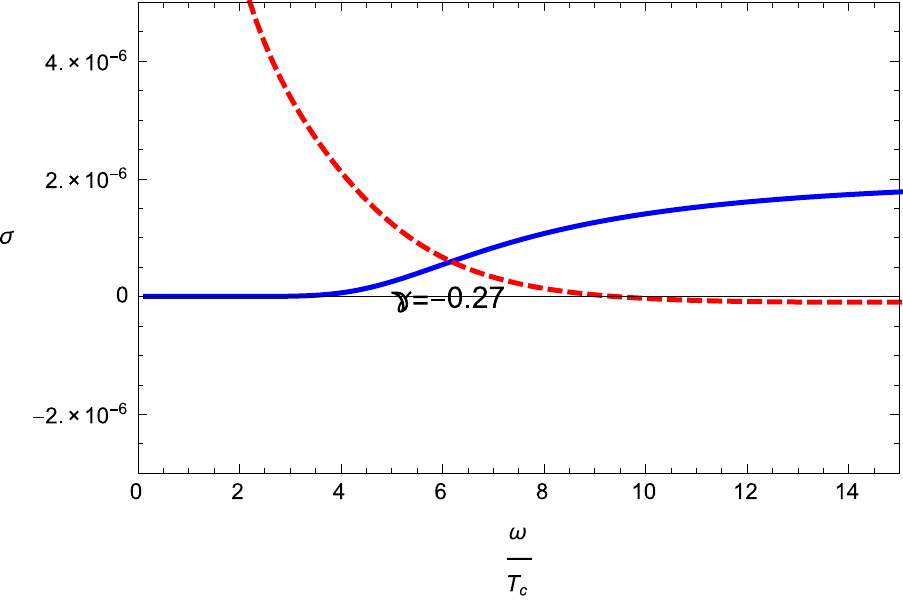}
\includegraphics[scale=.7]{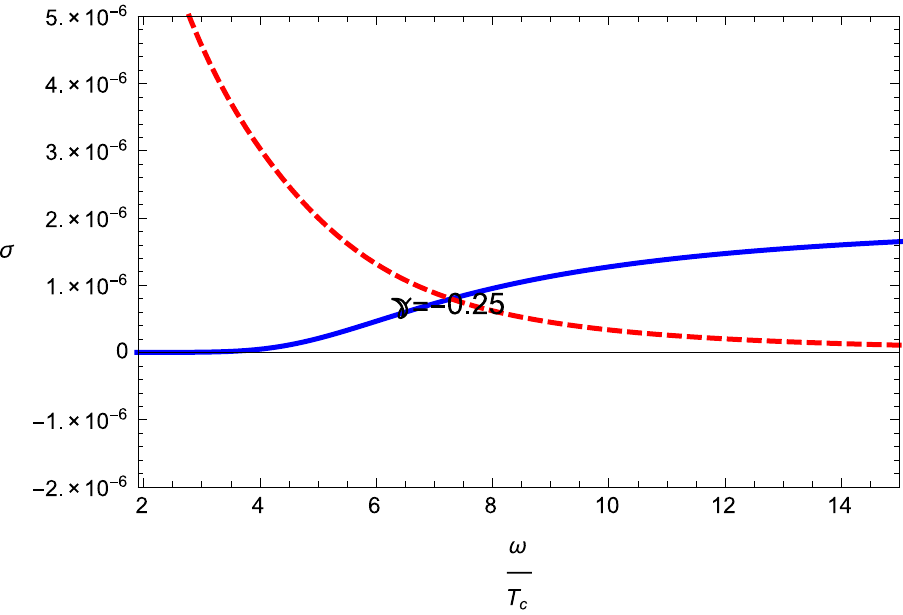}
\includegraphics[scale=.7]{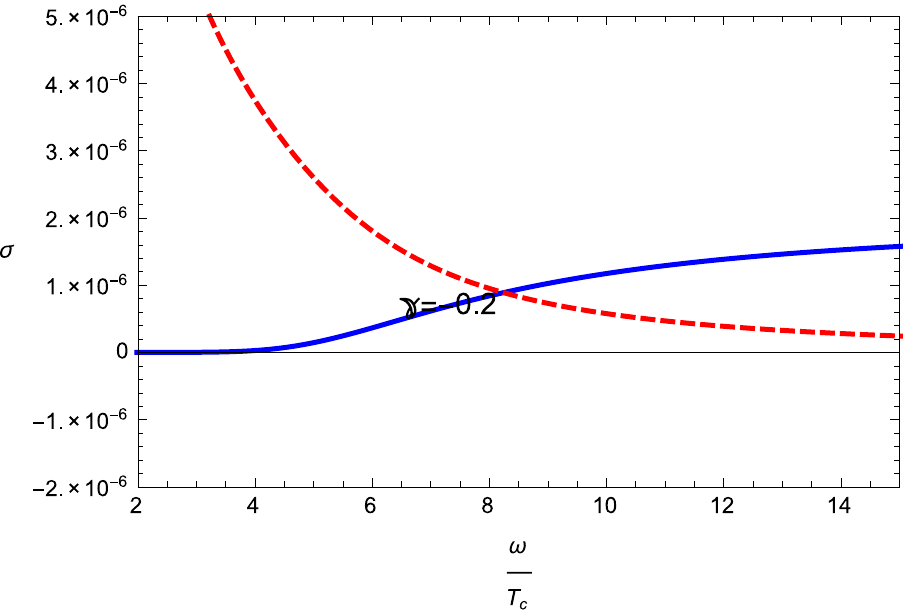}
\caption{The real (solid blue curve) and imaginary (dashed red curve) part of the AC conductivity versus frequency of the Weyl
model at $T/Tc \approx 0.0110567$ with $\Delta=2$, $z=2$, and $d=2$ for different $\gamma=-0.36,-0.33,-0.3,-0.27,-0.25,-0.2$ from top to bottom. }\label{fig8}
\end{figure}

In this section, we investigate the influence of the Weyl coupling, $\gamma$, and dynamical critical exponent, $z$, on the electrical conductivity. To calculate the electrical conductivity in the boundary field theory side, we need to consider the perturbation of the gauge field in the bulk. Therefore, we must add a small perturbation $\delta A=A_{y}(u)e^{-i\omega t} dy$ to the gauge field $A_{\mu}$ defined in the bulk geometry. The linearized equation of the perturbation $A_{y}$ turns out to be:
\begin{eqnarray}\label{eqE1}
&{{A}_{y}}^{''}&+A_{y}^{'}\left( \frac{{{U}_{2}}'}{{{U}_{2}}}+\frac{f'}{f}-\frac{d+z-3}{u} \right)+{{A}_{y}}\left( \frac{{{w}^{2}}}{{{f}^{2}}}{{u}^{2z-2}}-\frac{2{{\psi }^{2}}}{{{u}^{2}}f} \right)=0
\end{eqnarray}
Near the horizon $u \to 1$, we should consider the ingoing wave boundary condition for the electromagnetic field fluctuation in order to compute the retarded Green's function,
\begin{eqnarray}\label{form-27}
{{A}_{y}}(u)={{(1-u)}^{-\frac{i\omega }{4\pi T}}}(1+{{A}_{y1}}(1-u)+{{A}_{y2}}{{(1-u)}^{2}}+...).
\end{eqnarray}
Near the conformal boundary $u\rightarrow0$, the asymptotical expansion of $A_{y}(u)$ takes the following form:
\begin{equation}\label{form-29}
A_{y}(u)=A^{0}+A^{(d+z-2)}u^{d+z-2}+...
\end{equation}
Note that in the case of $z = 2$, $d = 2$, a logarithmic term $-A^{(0)} \omega^{2} u^{2}ln(\kappa u)$ should be added to the right hand side of \eqref{form-29}, where $\kappa$ is a constant.
According to the linear response theory, the conductivity is given by the following Kubo formula,
\begin{equation}\label{form-30}
\sigma (\omega)=\lim_{u\rightarrow 0}\frac{G^{R}(\omega, \overrightarrow{k}=0)}{i\omega}
\end{equation}
where, the retarded Green's function $G^{R}(\omega, \overrightarrow{k}=0)$ for the operator dual to gauge field can be computed through the recipe given in \cite{kobo1} (See appendix ~\ref{Appc}).
Therefore the electrical conductivity for our model can be obtained as follows:
\begin{equation}\label{form-32}
\sigma(\omega)=\frac{-i}{\omega}(d+z-2)\frac{A^{(d+z-2)}}{A^{(0)}}
\end{equation}
Now, we discuss the results for the conductivity obtained through the numerical solution of Eq. \eqref{eqE1}. The numerical results of the frequency dependent conductivity are illustrated in Figs. ~\ref{fig7} and ~\ref{fig8} for different values of $\gamma$ at $z=1,2$ in $d=2$. It should be noted that in $d=2$, the gamma bound is $-0.083<\gamma<0.083$ for $z=1$, and $\gamma<-0.19$ and $\gamma>0.38$ for $z=2$, respectively.

In each plot, the blue (solid) and red (dashed) lines represent the real part and imaginary part of the conductivity $\sigma(\omega)$, respectively. The imaginary part has a pole at $\omega=0$, which indicates that the real part contains a delta function according to the Kramers-Kronig relation \cite{Ref1, Ref2}. It is easy to show that there exists a gap in the conductivity, which rises quickly near the gap frequency $\omega_{g}$. The ratio of gap frequency over critical temperature $\omega/T_{c}$ is unstable and running with the Weyl coupling $\gamma$ in Fig.~\ref{fig7}. In other words, for z=1, the ratio $\omega_{g}/T_{c}$ increases with the fall of the Weyl parameter $\gamma$. In addition, the ratio $\omega_{g}/T_{c} > 8$ for all values of $\gamma$, and goes to 8 that is similar to the standard holographic superconductor model \cite{Dref1}, as we amplify the parameter  $\gamma$. This is in agreement with the cases in Gauss-Bonnet gravity \cite{Ref1}, in which $\omega_{g}/T_{c}$ is always greater than 8. On the other hand, Fig.~\ref{fig8} displays at $z=2$, the ratio $\omega_{g}/T_{c}<8$ for $\gamma<-0.19$. Especially, for $\gamma =- 0.36$, the value of ratio $\omega_{g}/T_{c} \approx 4$ which is different from the previous results. As a result, compared with the former models, it is interesting that $\omega_{g}/T_{c}$ in our model is closer to the weakly coupled BCS value of 3.5. 

Furthermore, in Fig.~\ref{fig8} a gap in the conductivity with a frequency $\omega_{g}$ becomes larger when we decrease the values of $\gamma$, while in Fig.~\ref{fig7} the the gap will be larger by increasing $\gamma$. For all cases considered here, we see from both figures that the real part of the conductivity is suppressed in the case of $z = 2$, compared to the case of $z = 1$. It shows the anisotropic effect of the background spacetime.
 In addition, we can clearly see from Fig.~\ref{fig8} that when $z=2$, the minimum of the imaginary part of the conductivity disappears, which means that the energy gap is no longer obvious.
\section{Summary and conclusions}\label{cap4}
The present paper sought to gain an  understanding of how the Weyl coupling, $\gamma$, and the Lifshitz scaling $z$ might affect the holographic superconductor. For this purpose the holographic superconductor model was constructed in the presence of Weyl corrections to the gravitational action in Lifshitz black-hole space-times.

Among the interesting results found  were the bounds on the Weyl coupling using certain constraints. These constraints were derived by considering that the causality is respected in the dual field theory on the boundary and that the energy flux is positive in the dual CFT analysis. In the logical range of Weyl coupling, we applied the matching method to study the effect of Lifshitz scaling on the Weyl holographic superconductor. In  the probe limit, the calculations showed that  critical temperature decreased with increasing $z$ for a fixed value of $\gamma$. This made condensation harder, while the critical temperature would be higher as we amplified the parameter, $\gamma$, for $z$ to be constant. The results were compared with those obtained from the numerical technique for $z=1$ ~\cite{ref6}.

Finally, the effect of an external static magnetic field on the Weyl model
of the holographic superconductor was investigated  by adding a magnetic field in the bulk. The results clearly revealed the dependence of the critical magnetic field on  parameters $\gamma$ and $z$. In this case, the height of the critical magnetic field, $B_{c}$, was found to  decrease with increasing  $z$. The critical magnetic field, $B_{c}$, was also observed to
  vanish faster for $ \gamma \neq 0 $  than for $\gamma=0$.

Finally, we calculated the conductivity of holographic superconductors numerically and find that the ratio $\omega/T_{c}$ is unstable and becomes larger when the Weyl coupling parameter $\gamma$ decreases at $z=1$. However, for $z=2$, the ratio $\omega/T_{c}$ will be smaller when the Weyl coupling parameter $\gamma$ decreases.
\section*{Acknowledgment}
We would like to thank department physics of Boston University for warm hospitality. This work is supported by Iranian National Science Foundation (INSF).   
\appendix
\section{Effective Potentials}\label{AppA}
The full wave functions (\ref{g8}) and (\ref{g3}) can now be rewritten  in the form of the Schrodinger
equation. By defining ${{A}_{t}}(u,\mathbf{q})={{G}_{1}}(u)\text{ }{{\Psi }_{1}}\text{ }(u,\mathbf{q})$ and making use of the coordinate transformation, ${s}'={{{u}^{z-1}}}/{f}$, one could rewrite Eq. \eqref{g8} in the form of  the Schr\"{o}dinger equation as follows:
\begin{eqnarray}\label{g13}
\partial _{s}^{2}{{\Psi }_{1}}(s)+V(s){{\Psi }_{1}}(s)={{\omega }^{2}}~{{\Psi }_{1}}(s)
\end{eqnarray}
where,  $G_{1}(s)$ satisfies the following equation:
\begin{eqnarray}\label{g12}
{{\partial }_{s}}{{G}_{1}}-\left[ \frac{{s}''+{{H}_{1}}{s}'}{2{{{{s}'}}^{2}}} \right]\text{ }{{G}_{1}}=0
\end{eqnarray}
It is easy to introduce the effective potential $V (s)$ in Eq. \eqref{g13} in the $u$ coordinate as follows:
\begin{eqnarray}\label{g14}
V\left( u \right)={{V}_{0}}\left( u \right){{q}^{2}}+{{V}_{1}}\left( u \right)
\end{eqnarray}
where,
\begin{eqnarray}\label{g15}
 {{V}_{1}}&=&\frac{{{f}^{2}}{{u}^{2-2z}}}{4}\left[ \frac{{{{{U}'}}_{2}}}{{{U}_{2}}}\left[ \frac{3{{{{U}'}}_{2}}}{{{U}_{2}}}-\frac{2{f}'}{f} \right]-\frac{2{{{{U}''}}_{2}}}{{{U}_{2}}} \right]+\frac{{{f}^{2}}{{u}^{1-2z}}}{2}\left[ \frac{{{z}_{1}}{{{{U}'}}_{2}}}{{{U}_{2}}}+\frac{{f}'\left( d-2 \right)}{f} \right] \\
 \nonumber & +&\frac{{{f}^{2}}{{u}^{-2z}}}{2}\left[ {{z}_{1}}\left( -d+2 \right)+d\left( 1-d \right) \right] \\
{{V}_{0}}&=&-\frac{f{{U}_{2}}{{u}^{2-2z}}}{{{U}_{1}}}\label{g16}
\end{eqnarray}
On the other hand, we can repeat  similar algebraic calculations for the transverse vector mode satisfying Eq. \eqref{g3} by writing  ${{A}_{y}}\left( u \right)={{G}_{2}}\left( u \right){{\Psi}_{2}}\left( u , \textbf{q}\right)$. Therefore, we have:
\begin{equation}\label{g19}
\partial _{s}^{2}{{\Psi }_{2}}(s)+W(s){{\Psi }_{2}}(s)={{\omega }^{2}}{{\Psi }_{2}}(s)
\end{equation}
where the effective potential, $W$, is defined as:
\begin{eqnarray}\label{g20}
W\left( u \right)={{W}_{0}}\left( u \right){{q}^{2}}+{{W}_{1}}\left( u \right)
\end{eqnarray}
with
\begin{eqnarray}\label{g21}
   {{W}_{1}}&=&\frac{{{f}^{2}}{{u}^{2-2z}}}{4}\left[ \frac{2{{{{U}''}}_{2}}}{{{U}_{2}}}-\frac{{{{{U}'}}_{2}}}{{{U}_{2}}}\left[ \frac{{{{{U}'}}_{2}}}{{{U}_{2}}}-\frac{2{f}'}{f} \right] \right]+\frac{{{f}^{2}}{{u}^{1-2z}}}{2}\Big[ 2\left( d+z-3 \right)\frac{{{{{U}'}}_{2}}}{{{U}_{2}}}\\
\nonumber &+&\left( d+2z-4 \right)\frac{{{f}'}}{f} \Big]   + \frac{{{f}^{2}}{{u}^{-2z}}}{4}\left( d+2z-4 \right)\left( d-4 \right) \\
{{W}_{0}}&=&\frac{f{{U}_{3}}{{u}^{2-2z}}}{{{U}_{2}}}\label{g22}
\end{eqnarray}
where $z_2=4-2d-z_{1}$. Moreover, the function $G_{2}$  has to satisfy the following relation:
\begin{eqnarray}\label{g12}
\partial _{s}G_{2}-\left[ \frac{{{s}^{''}}+{K}s^{'}}{2{{s^{'}}^{2}}} \right]G_{2}=0
\end{eqnarray}
where, $K=\left[ \frac{{{U}_{2}}'}{{{U}_{2}}}+\frac{f'}{f}-\frac{z+d-3}{u} \right]$.
\section{Potential Expansion}\label{AppB}
One can expand the effective potential $V_{0}\left( u \right)$ near the boundary as follows:
\begin{equation}\label{rq1}
{{V}_{0}}\simeq \sum\limits_{n=0}^{\infty }{\sum\limits_{a+b+c=n}{-\frac{{{u}^{2-2z+n}}}{a!b!c!}\left[ {{\left. {{f}^{\left( a \right)}}{{U}_{2}}^{\left( b \right)}{{\left( \frac{1}{{{U}_{1}}} \right)}^{\left( c \right)}} \right|}_{u=0}} \right]}}
\end{equation}
where, $a$, $b$, and $c$ may  consist of the fractional derivative. According to Eq. \eqref{ssq1}, the highest degree of  polynomial $U_{1}$ and $U_{2}$ is ${d+z}$. Therefore, the non-vanishing terms of the above expansion can generally be written as follows:
\begin{equation}\label{ss1}
{{V}_{0}}\simeq - {{\left. \frac{{{U}_{2}}f}{{{U}_{1}}} \right|}_{u=0}}{{u}^{2-2z}}-\frac{1}{(d+z)!}{{\left. \left( \frac{{{f}^{\left( d+z \right)}}{{U}_{2}}}{{{U}_{1}}}+\frac{fU_{2}^{\left( d+z \right)}}{{{U}_{1}}}+\frac{f{{U}_{2}}U_{1}^{\left( d+z \right)}}{U_{1}^{2}} \right) \right|}_{u=0}}{{u}^{d+2-z}}
\end{equation}
In the following, we obtain ranges of the Weyl coupling in the three cases of $0<z<1$, $z=1$, and $z>1$ based on the causality requirement \eqref{QEQ2}.  It is obvious that all the powers of $u$ will be positive for $0<z<1$. So, there is no constraint on the Weyl coupling at $u \to 0$. Moreover, for $z>1$, the first order of expansion is divergent at the boundary. In order to satisfy the causality at the boundary, the first term must be positive. It forces the following condition on the Weyl coupling range:
\begin{equation}
\gamma >\gamma_{2}=\frac{d(d+1)}{8z(d-1)(z-1)}
\end{equation}
When we consider $z=1$, however, the first term will be identity and the other terms can be important. From the causality constraint, i.e., $V_{0}(u)<1$, the second term in Eq. \eqref{ss1} must be negative, i.e.,
\begin{equation}
\gamma<\gamma_{1}=\frac{1}{4(d+1)(d-1)}
\end{equation}
In a similar fashion one can get the following  non-equality expressions from the ${{W}_{0}}(u)$ expansion near the boundary by substituting $\left\{ {{U}_{1}},{{U}_{2}} \right\}$ with $\left\{ {{U}_{2}},{{U}_{3}} \right\}$.
\begin{equation}
\begin{array}{ccc}
   \gamma >{{\gamma }_{4}}=\frac{-1}{4(d+1)} & if & z=1 \\
& & \\
  \gamma <{{\gamma }_{3}}=\frac{-d(d+1)}{16z(z-1)} & if & z>1
 \end{array}
\end{equation}
For $0<z<1$, the results of the intersection of the bounds show that in the case of $0<z<1$, there is no constraint on the Weyl coupling. Moreover, the gamma bound for $\text{z}>\text{1}$ is given by $\gamma >{{\gamma }_{2}}$ and $ \gamma <{{\gamma }_{3}}$, while for $z=1$, the bound is $\gamma_{4}<\gamma<\gamma_{1}$.
We also consider another constraint on the expansions of ${{V}_{0}}(u)$ and ${{W}_{0}}(u)$ near the horizon to produce a positive energy in all directions for a consistent CFT. The expansion of ${{V}_{0}}(u)$ near the horizon is as follows:
\begin{equation}
{{V}_{0}}\simeq- {{\left. \frac{{{U}_{2}}f}{{{U}_{1}}} \right|}_{u=1}}-{{\left. \left[ \frac{{{{{U}'}}_{2}}f}{{{U}_{1}}}+\left( {f}'-2(z-1)f \right)\frac{{{U}_{2}}}{{{U}_{1}}}-\frac{{{U}_{2}}{{{{U}'}}_{1}}f}{{{U}_{1}}} \right] \right|}_{u=1}}(u-1)+...
\end{equation}
Near the horizon, due to $f(1)=0$, the first term vanishes and the second term always vanishes at $u = 1$; thus, we can obtain the limitation on the Weyl coupling immediately near the horizon, indicating the presence of the negative potential  there.
Therefore, one needs to take $U_{2}(1)/U_{1}(1)<0$ in $V_{0}$ and $U_{3}(1)/U_{2}(1)>0$ in $W_{0}$ to obtain another range of the Weyl coupling.
\section{The Retarded Green's function}\label{Appc} 
 Using the AdS/CFT correspondence and following prescription given in Ref.  \cite{kobo1}, we can calculate the retarded Green's function. The action of gauge field with the Weyl correction is,
\begin{eqnarray}\label{form-33}
S&=&\int d^{d+2}y \sqrt{-g}\Big[\frac{1}{4}\nabla_{\mu} F^{\mu\nu}A_{\nu}-\frac{1}{4}\nabla_{\mu} (F^{\mu\nu}A_{\nu})\nonumber\\
&-&\frac{1}{4}\nabla_{\mu} F^{\mu\nu}A_{\mu}+\frac{1}{4}\nabla_{\nu}(F^{\mu\nu}A_{\mu})\nonumber\\
&+&\gamma \nabla_{\mu}(C^{\mu\nu\rho\sigma}A_{\nu}F_{\rho\sigma})-\gamma \nabla_{\nu}(C^{\mu\nu\rho\sigma}F_{\rho\sigma})A_{\nu}\nonumber\\
&-&\gamma \nabla_{\nu}(C^{\mu\nu\rho\sigma}A_{\mu}F_{\rho\sigma})+\gamma \nabla_{\nu}(C^{\mu\nu\rho\sigma}F_{\rho\sigma})A_{\mu}\Big]\nonumber\\
&=&-\int d^{d+2}y \sqrt{-g}\Big[\frac{1}{2}\nabla_{\mu} (F^{\mu\nu})A_{\nu}-2\gamma \nabla_{\mu}(C^{\mu\nu\rho\sigma}F_{\rho\sigma}A_{\nu})\Big]\nonumber\\
&+&\int d^{d+2}y \sqrt{-g}\Big[\frac{1}{2}\nabla_{\mu} (F^{\mu\nu})A_{\nu}-2\gamma \nabla_{\mu}(C^{\mu\nu\rho\sigma}F_{\rho\sigma})A_{\nu}\Big]\nonumber\\
&=&-\int d^{d+2}y \sqrt{-g}\Big[\frac{1}{2}\nabla_{\mu} (F^{\mu\nu}A_{\nu})-2\gamma \nabla_{\mu}(C^{\mu\nu\rho\sigma}F_{\rho\sigma}A_{\nu})\Big]\nonumber\\
&=&-\int_{\partial M}d^{d+1}y\sqrt{-h}\Big[\frac{1}{2}F^{\mu\nu}n_{\mu}A_{\nu}-2\gamma C^{\mu\nu\rho\sigma}F_{\rho\sigma}n_{\mu}A_{\nu}\Big]
\end{eqnarray}
the action reduces to a surface term due to the bulk contribution vanishing so we can obtain the action as follows
\begin{eqnarray}\label{form-34}
S&=&-\frac{1}{2}\int_{\partial M}d^{d+1}y\sqrt{-h}g^{uu}g^{yy}(1-8\gamma g^{uu}g^{yy}C_{uyuy})n_{u}A_{y}\partial_{u}A_{y}\nonumber\\
&=&\frac{1}{2}\int_{\partial M}d^{d+1}y\frac{1}{u^{d+z-1}}f(u)\Big(1+8\gamma \Big (\frac{d-1}{2}\Big)\xi (u)\Big) A_{y}\partial_{u}A_{y}
\end{eqnarray}
Near the boundary $(u\rightarrow 0)$, we may neglect the Weyl correction and the action can be written as follows
\begin{equation}\label{form-36}
S=-\frac{1}{2}\int_{\partial M}d^{d+1}y \frac{1}{u^{d+z-1}}f(u)A_{y}\partial_{u}A_{y}|_{u\rightarrow 0}
\end{equation}
For the standard AdS/CFT correspondence we have
\begin{equation}\label{form-37}
S=\frac{1}{2}\int_{\partial M}d^{d+1}y\frac{1}{2\pi ^{4}}A_{y}(-k)G^{R}(k)A_{y}(k)|{u\rightarrow 0}
\end{equation}
by comparing (\ref{form-36}) and (\ref{form-37}) yields:
\begin{equation}
G^{R}(\omega, k^{\rightarrow}=0)=f(u)\frac{1}{u^{d+z-1}}\frac{A_{y}(u,k)\partial_{u}A_{y}(u,k)}{A_{y}(u,-k)A_{y}(u,k)}
\end{equation}
Therefore we find that the Weyl correction has no effect on reterded Green's function and is the same as the Einstein theory for the standard Maxwell field.




\end{document}